\documentclass[prb,twocolumn,showpacs,preprintnumbers,amsmath,amssymb]{revtex4}

\usepackage{amsmath,amssymb,amsfonts} 	
\usepackage[dvips]{graphicx}
\usepackage[latin1]{inputenc}
\usepackage{subfigure}
\begin{document}

\title{The random phase approximation applied to solids, molecules, and graphene-metal interfaces: From weak to strong binding regimes}
\author{Thomas Olsen}
\email{tolsen@fysik.dtu.dk}
\author{Kristian S. Thygesen}

\affiliation{Center for Atomic-Scale Materials Design (CAMD) and Center for Nanostructured Graphene (CNG),
	     Department of Physics, Technical University of Denmark,
	     DK--2800 Kongens Lyngby, Denmark}

\date{\today}

\begin{abstract}
The random phase approximation (RPA) is attracting renewed interest as a universal and accurate method for first-principles total energy calculations. The RPA naturally accounts for long-range dispersive forces without compromising accuracy for short range interactions making the RPA superior to semi-local and hybrid functionals in systems dominated by weak van der Waals or mixed covalent-dispersive interactions. In this work we present plane wave-based RPA calculations for a broad collection of systems with bond types ranging from strong covalent to van der Waals. Our main result is the RPA potential energy surfaces of graphene on the Cu(111), Ni(111), Co(0001), Pd(111), Pt(111), Ag(111), Au(111), and Al(111) metal surfaces, which represent archetypical examples of metal-organic interfaces. Comparison with semi-local density approximations and a non-local van der Waals functional show that only the RPA captures both the weak covalent and dispersive forces which are equally important for these systems. We benchmark our implementation in the GPAW electronic structure code by calculating cohesive energies of graphite and a range of covalently bonded solids and molecules as well as the dissociation curves of H2 and H2+. These results show that RPA with orbitals from the local density approximation suffers from delocalization errors and systematically underestimates covalent bond energies yielding similar or lower accuracy than the Perdew-Burke-Ernzerhof (PBE) functional for molecules and solids, respectively.
\end{abstract}
\maketitle

\section{Introduction}
The RPA was first introduced by Bohm and Pines \cite{bohm1,bohm2,bohm3} more than 60 years ago and thus predates both Kohn-Sham Density Functional Theory (DFT) and the formal developments in many-body perturbation theory.\cite{gellmann}  A most important property of the RPA is the explicit incorporation of screening in correlated quantities. The screening allows one to treat the electrons in metals as nearly independent "quasi-particles" interacting through a screened effective Coulomb interaction and explains why the single-particle picture often gives a decent description of solids, despite the large strength of the bare Coulomb interaction.\cite{landau} The RPA takes screening into account by summing a certain class of Feynman diagrams to infinite order in the Coulomb interaction and this allows one to evaluate correlation energies of metallic systems, which diverge in perturbative treatments.\cite{flensberg}

In the context of DFT, the correlation energy can be expressed in terms of the interacting response function using the adiabatic-connection and fluctuation-dissipation theorem (ACFD).\cite{langreth} Time-dependent DFT relates the interacting response function to the Kohn-Sham response function through an exchange-correlation kernel and RPA is then the simplest possible approximation where the kernel is neglected all together.\cite{marques} In principle, it is possible to calculate a local RPA potential using the optimized effective potential method, and solve the resulting Kohn-Sham equations for a self-consistent RPA density and total energy. However, the huge computational cost of such calculations has so far limited the self-consistent approach to atoms\cite{hellgren1,hellgren2} and simple molecules.\cite{hellgren3} Instead, it is often assumed that the effect of selfconsistency is of minor concern and RPA calculations are performed non-selfconsistently using Kohn-Sham or Hartree-Fock eigenstates and eigenenergies. The approach then becomes equivalent to RPA from perturbation theory. 

The use of RPA as a tool for \textit{ab initio} total energy calculations was pioneered by Furche\cite{furche,eshuis} who calculated the RPA atomization energies of small molecules and found that RPA has a general tendency to underbind. It was also demonstrated that the atomic limit of N$_2$ dissociation was reproduced by RPA if the reference is taken with respect to the RPA energy of two isolated N atoms. Subsequently, the performance of RPA has been examined systematically for molecular dissociation\cite{paier, ren_review}, cohesive energies of solids,\cite{harl08,harl09,harl10} surface properties and adsorbates,\cite{schimka} barrier heights,\cite{paier, ren_review} ionization potentials\cite{eshuis} van der Waals bonded dimers,\cite{ren_review} and van der Waals bonded two-dimensional materials.\cite{marini,lebegue,sachs} Compared to DFT calculations, the computational load of RPA calculations can represent a significant barrier for applications to large electronic systems. Nevertheless, due to increasing access to high performance computational resources, there is a rapidly growing interest in the method and RPA is now slowly emerging as a standard tool in the electronic structure community.

In general, RPA seems to be inferior to Generalized Gradient Approximations (GGA) and hybrid exchange-correlation functionals for the description of covalent bonds. However, the non-local nature of RPA makes its superior to any semi-local or hybrid functional for dispersive interactions. Several attempts have been made to construct effective non-local van der Waals functionals, which capture long range dispersive interactions and are comparable to GGA calculations in computational requirements.\cite{andersson, dobson_wang99,dion,vydrov1,vydrov2} In many cases, such functionals have been successful, but the approach is rather sensitive to the choice of exchange kernel and typically fails to give a qualitative description if both covalent and dispersive interactions are important.\cite{langreth09,wellendorff} In contrast, the RPA correlation energy is naturally combined with exact exchange, does not rely on error cancellation or any fitted parameters, and are able to describe intricate bonds with mixed covalent and dispersive character.\cite{olsen1,mittendorfer} For an accurate description of both strong covalent bonds \textit{and} dispersive interactions, it is necessary to apply beyond-RPA methods such as screened second order exchange,\cite{gruneis} time-dependent exact exchange,\cite{hesselmann} or renormalized adiabatic kernels.\cite{olsen2} However, such developments are outside the scope of the present paper and we will focus on RPA in the following.

Two-dimensional layered compounds such as graphite, hexagonal boron nitride, and transition metal dichalcogenides, constitute a particular class of materials where it is vital to incorporate dispersive interactions in order to obtain a quantitative description of the bulk properties. RPA has been shown to provide an accurate description of van der Waals bonds in these systems\cite{marini,lebegue,bjorkman}, whereas semi-local and van der Waals functionals can give rise to qualitatively wrong results. Moreover, the discovery and characterization of isolated graphene sheets,\cite{novoselov1,geim} has triggered a vast amount of research in this intriguing two-dimensional material. In particular, graphene shows a remarkably high intrinsic carrier mobility, and therefore seems very well suited for nanoscale electronics devices. For such applications, the coupling to metal contacts plays a fundamental role and measurements show that graphene binds very differently on various metal surfaces. Understanding the interactions between graphene and metal surfaces \cite{batzill}, therefore becomes a most important task since the adsorption geometry and bond distance may have drastic consequences for the electronic structure and transport properties of adsorbed graphene layers. For example, experiments have demonstrated that Pd(111), Co(0001), and Ni(111) induce a band gap in adsorbed graphene sheets, which signals a covalent bond with the metal \cite{eom,kwon,varykhalov}. In contrast, adsorption on Cu(111), Ag(111), Au(111), and Pt(111) do not change the graphene band structure significantly \cite{sutter,shikin,klusek}. On the theoretical side, most studies have been limited to semi-local approximations\cite{kelly} and van der Waals functionals\cite{vanin, hamada}. While some agreement with experiment was obtained in Ref. [\onlinecite{hamada}] for a certain van der Waals functional, the large discrepancy between various choices of functional is clearly unsatisfactory.

Here we apply RPA to calculate the binding energy curves of graphene on Cu(111), Ni(111), Co(0001), Pd(111), Pt(111), Ag(111), Au(111), and Al(111) metal surfaces. The results for Cu(111), Ni(111), and Co(0001) have been obtained previously\cite{olsen1,mittendorfer} but are reproduced here for completeness. We also show that the slight discrepancy between the RPA curve for graphene on Ni(111) in Refs. [\onlinecite{olsen1}] and [\onlinecite{mittendorfer}] were caused by insufficient $k$-point sampling in Ref. [\onlinecite{olsen1}]. For all the metal surfaces except Pd(111), we find good agreement with experiment. The deviation in the case of Pd, is most likely related to the large discrepancy between the metal and graphene unit cells and a proper Moire structure is needed in order to compare with experiments in this case.

The paper is organized as follows. In section \ref{method} we outline the general method used to obtain RPA total energies and present details on the plane wave implementation applied in the present work. In section \ref{results}, the RPA potential energy curves for graphene adsorbed on 8 different metal surfaces are presented and compared with semi-local approximations for the exchange-correlation energy and a standard van der Waals functional.
We then assess the quality and wide applicability of the method and implementation by benchmarking calculated results for dissociation of graphite, properties of bulk solid state systems and molecular atomization energies. In appendices \ref{Ni_convergence} and \ref{CO_convergence}, we present detailed convergence tests for the RPA potential energy curves of graphene on Ni(111) and for the atomization energy of the CO molecule.

\section{Method}\label{method}
\subsection{Theory}
Using the adiabatic connection and fluctuation-dissipation theorem (ACFD), the exchange-correlation energy can be written as:
\begin{align}\label{E_xc}
 E_{xc}=-\int_0^1d\lambda\int_0^\infty\frac{d\omega}{2\pi}\text{Tr}\Big\{v[\tilde{n}2\pi\delta(\omega)+\chi^{\lambda}(i\omega)]\Big\},
\end{align}
where $\tilde{n}(\mathbf{r},\mathbf{r}')=n(\mathbf{r})\delta(\mathbf{r}-\mathbf{r}')$ and $v$ is the Coulomb interaction. Here $n(\mathbf{r})$ is the density, which by definition is constant along the adiabatic connection and $\chi^\lambda(i\omega)$ is the interacting response function of a system with $v\rightarrow\lambda v$ evaluated at imaginary frequencies. It is standard practice to divide $E_{xc}$ into an exchange part $E_x$ obtained by setting $\lambda=0$ in the integrand and a correlation part $E_c$, which is the remainder. One then obtains 
\begin{align}
 E_{x}&=-\int_0^\infty\frac{d\omega}{2\pi}\text{Tr}\Big\{v[\tilde{n}2\pi\delta(\omega)+\chi^{KS}(i\omega)]\Big\},\label{E_x}\\
E_{c}&=-\int_0^1d\lambda\int_0^\infty\frac{d\omega}{2\pi}\text{Tr}\Big\{v[\chi^\lambda(i\omega)-\chi^{KS}(i\omega)]\Big\},\label{E_c}
\end{align}
where $\chi^{KS}(i\omega)$ is the response function of the non-interacting Kohn-Sham system. A major advantage of this separation is that the exchange energy can be evaluated exactly and one only needs to approximate $\chi^\lambda$ to obtain $E_c$.

The Random phase approximation for the interacting response function can be derived in several ways, but in the present context it is convenient to use time-dependent density functional theory, from which it is it is straightforward to show that
\begin{align}\label{TDDFT}
\chi^\lambda(i\omega)=\chi^{KS}(i\omega)+\chi^{KS}(i\omega)\big[\lambda v+f^\lambda_{xc}(i\omega)\big]\chi^\lambda(i\omega),
\end{align}
where $f_{xc}^\lambda$ is the exchange-correlation kernel. The RPA is then obtained by taking $f_{xc}^\lambda=0$ and one is left with
\begin{align}\label{chi_rpa}
\chi^\lambda_{RPA}(i\omega)=\big[1-\chi^{KS}(i\omega)\lambda v\big]^{-1}\chi^{KS}(i\omega).
\end{align}
Inserting this into the expression for the correlation energy and carrying out the coupling constant integration yields
\begin{align}\label{RPA}
 E_c^{RPA}=&\int_0^\infty\frac{d\omega}{2\pi}\text{Tr}\Big\{\ln[1-v\chi^{KS}(i\omega)] +v\chi^{KS}(i\omega)\Big\}.
\end{align}
For spin polarized systems the correlation energy involves the spin summed response function $\tilde \chi^\lambda=\sum_{\sigma\sigma'}\chi^\lambda_{\sigma\sigma'}$. Using that $v$ is independent of spin, it is straightforward to show that $\tilde \chi^\lambda$ satisfies Eq. \eqref{chi_rpa} if $\chi^{KS}$ replaced by $\chi^{KS}_\downarrow+\chi^{KS}_\uparrow$. This would not be true if a spin-dependent $f_{xc}$ were included in Eq. \eqref{TDDFT} and comprises a major simplification of RPA calculations involving spin polarized systems. 

\subsection{Plane wave implementation}
For solid state systems it is convenient to evaluate the response function in a plane wave representation. The number of plane waves required at a given energy cutoff scales as $N_G\sim V_{cell}$, which determines the dimension of the response function. For isolated atoms and molecules where large unit cells is needed in order to decouple periodic images, the response function may become prohibitly large and the implementation is not well suited for large molecular systems. However, as will be shown below, it is possible to calculate atomization energies for small molecules although the computational time is much larger than implementations utilizing atomic basis sets.

In a plane wave basis the Kohn-Sham response function is 
\begin{align}\label{response}
\chi^{KS}_{\mathbf{G}\mathbf{G}'}(\mathbf{q},i\omega)&=\frac{1}{V}\sum_{\mathbf{k}\in BZ}\sum_{n,n'}\frac{f_{n\mathbf{k}}-f_{n'\mathbf{k}+\mathbf{q}}}{i\omega+\varepsilon_{n\mathbf{k}}-\varepsilon_{n'\mathbf{k}+\mathbf{q}}}\\
\times\langle\psi_{n\mathbf{k}}|&e^{-i(\mathbf{q}+\mathbf{G})\cdot\mathbf{r}}|\psi_{n'\mathbf{k}+\mathbf{q}}\rangle\langle\psi_{n'\mathbf{k}+\mathbf{q}}|e^{i(\mathbf{q}+\mathbf{G}')\cdot\mathbf{r}}|\psi_{n\mathbf{k}}\rangle\notag,
\end{align}
where $\varepsilon_{n\mathbf{k}}$ are Kohn-Sham eigenvalues, $f_{n\mathbf{k}}$ are occupation numbers, and $|\psi_{n\mathbf{k}}\rangle$ are the Kohn-Sham eigenstates normalized in the unit cell with volume $V$. The trace in Eq. \eqref{RPA} then becomes a trace over plane waves and a Brillouin Zone integral over $\mathbf{q}$ which is sampled on a uniform grid:
\begin{align}\label{RPA_PW}
&E_c^{RPA}=\int_0^\infty\frac{d\omega}{2\pi}\frac{1}{N_{\mathbf{q}}}\sum_{\mathbf{q}\in BZ}\\
&\times\text{Tr}\Big\{\ln[1-v(\mathbf{q})\chi^{KS}(\mathbf{q},i\omega)] +v(\mathbf{q})\chi^{KS}(\mathbf{q},i\omega)\Big\}.\notag
\end{align}
The plane wave representation of the coulomb interaction is $v:=4\pi\delta_{\mathbf{G}\mathbf{G}'}/|\mathbf{q}+\mathbf{G}|^2$ and the trace of the logarithm is most easily evaluated by using that $\text{Tr}[\ln(A)]=\ln[\det(A)]$. The exact exchange energy (EXX) Eq. \eqref{E_x} becomes
\begin{align}\label{EXX_PW}
E_x^{EXX}=&-\frac{1}{N_{\mathbf{q}}N_{\mathbf{k}}}\sum_{n,n'}\sum_{\mathbf{k},\mathbf{q}\in BZ}f_{n\mathbf{k}}\theta(\varepsilon_{n\mathbf{k}}-\varepsilon_{n'\mathbf{k+q}})\notag\\
&\times\sum_{\mathbf{G}} v_\mathbf{G}(\mathbf{q})|\langle\psi_{n\mathbf{k}}|e^{-i(\mathbf{q}+\mathbf{G})\cdot\mathbf{r}}|\psi_{n'\mathbf{k}+\mathbf{q}}\rangle|^2
\end{align}
The expression is derived from the ACFD and differs from the standard expression for exact exchange energy, if the occupation numbers are not integer valued. However, as discussed in Ref. [\onlinecite{harl10}], it is natural to apply Eq. \eqref{EXX_PW} when the exact exchange energy is combined with the RPA correlation energy. For metals, it is customary to aid convergence by smearing the occupation factors by an artificial electronic temperature and it has been shown empirically that Eq. \eqref{EXX_PW} is less sensitive to the width of the artificial smearing function $f_{n\mathbf{k}}$ than the standard expression for exact exchange.

The $\mathbf{q}=0$ terms in Eqs. \eqref{RPA_PW} and \eqref{EXX_PW} require a special treatment since $v(\mathbf{q})$ diverges as $\mathbf{q}\rightarrow0$ when $\mathbf{G}=0$. The divergence is, however, integrable and the terms yield a finite contribution. For the exact exchange part we apply the method of Gygy and Baldereschi \cite{gygi, sorouri} where the Coulomb interaction is multiplied by a Gaussian regularization and the $\mathbf{q}=0$ term can be integrated analytically in the limit of infinitely dense $k$-point sampling. The correlation energy is evaluated by replacing $|\psi_{n\mathbf{k+q}}\rangle$ by its first order perturbative expansion in the $\mathbf{k}\cdot\mathbf{p}$ Bloch Hamiltonian. For $n\neq n'$ this yields  
\begin{align}\label{optical}
\langle\psi_{n\mathbf{k}}|e^{-i\mathbf{q}\cdot\mathbf{r}}|\psi_{n'\mathbf{k}+\mathbf{q}}\rangle_{\mathbf{q}\rightarrow0}=\frac{\langle\psi_{n\mathbf{k}}|-i\mathbf{q}\cdot\mathbf{\nabla}|\psi_{n'\mathbf{k}}\rangle}{\varepsilon_{n'\mathbf{k}}-\varepsilon_{n\mathbf{k}}}
\end{align}
and the $\mathbf{q}$ on the right hand side cancels the diverging Coulomb interaction in Eq. \eqref{RPA_PW}. The limit clearly depends on the polarization of $\mathbf{q}$ and for non-isotropic systems we evaluate the $\mathbf{q}=0$ contribution to the correlation energy by averaging over non-equivalent polarizations. The method may fail for systems with a high density of states near the Fermi level such as certain transition metals, since the denominator in Eq. \eqref{optical} approach zero for low energy transitions. In principle, the problem should be solved by using degenerate perturbation theory, but there is no unique way of defining which states to treat as degenerate for a given $\mathbf{k}$. Instead we take a pragmatic point of view and simply exclude the $\mathbf{q}=0$ term in the evaluation of \eqref{RPA_PW} and \eqref{EXX_PW} for systems involving transition metals. This procedure has been shown to exhibit fast convergence for $E_x+E_c$ with respect to $k$-point sampling \cite{harl10}, whereas the individual exchange and correlation terms converge rather slowly when $\mathbf{q}=0$ is excluded. 

For all calculations we use Gamma-centered uniform $k$-point grids since the $k$-points and $q$-points then coincide. This ensures a much more efficient symmetry reduction of the $q$-points than if a shifted $k$-point grid were to be used. To evaluate the response function we usually choose a cutoff energy $E_{cut}^\chi$, which is smaller than the cutoff used to obtain the input eigenstates and eigenenergies and the set of included plane waves are determined by $|\mathbf{q}+\mathbf{G}|^2/2<E_{cut}^\chi$. The dimension of the response function $\chi_{\mathbf{G}\mathbf{G}'}(\mathbf{q})$ thus depends on $\mathbf{q}$, which ensures a smooth dependence on the cutoff energy for periodic systems. In all calculations, the number of bands used to evaluate the response function is set equal to the number of plane waves determined by the cutoff. This approach is appealing, since the RPA calculations then only depends on a single convergence parameter. Furthermore, it will be more straightforward to compare with properties of the homogeneous electron gas and the Lindhard function, since in that case, the eigenstates coincide with plane waves.

The response function is evaluated at the imaginary frequency axis, where it varies rather smoothly and allows for an efficient numerical integration. Typically, the density of states near the Fermi level determines how much structure the response function exhibits near $\omega=0$. The frequency integration in Eq. \eqref{RPA_PW} is carried out using 16 Gauss-Legendre points with a weight function ensuring that the integral of $f(x)\propto x^{(1/B-1)}\exp{[-\alpha x^{1/B}]}$ is reproduced exactly. Here $\alpha$ is determined by the highest frequency point, which we position at $800\;eV$ for all calculations. $B$ determines the density of frequency points close to $\omega=0$ and we use $B=2.0$ for systems with a gap and $B=2.5$ for metals. With this frequency sampling the RPA correlation energies are converged to within a few meV. 

Since, the present approach is not self-consistent, one has to choose a set of orbitals, on which Eqs. \eqref{RPA_PW}-\eqref{EXX_PW} are evaluated. We have compared the RPA potential energy surface for graphene on Ni(111) using self-consistent Perdew-Burke-Ernzerhof (PBE) orbitals with the result obtained with self-consistent Local Density-Approximation (LDA) orbitals and the results are very similar. In other cases, however, there may be a significant dependence on the initial orbitals. For example, the RPA atomization energy of O$_2$ and CO was shown to differ by $\sim0.4$ eV when comparing LDA and PBE initial orbitals \cite{olsen2}. Even larger differences have been observed when comparing Hartree-Fock (HF) and PBE initial orbitals\cite{ren} and the most accurate RPA results are obtained when combining self-consistent EXX with RPA correlation energies evaluated on PBE or similar initial orbitals. Alternatively, the effect of non-selfconsistency can be corrected by including single-excitation contributions to the correlation energy \cite{ren}. Unless stated otherwise, all RPA calculations below are performed with PBE Kohn-Sham orbitals and eigenvalues. The non-selfconsistent EXX will be referred to as Hartree-Fock (HF).

The response function, EXX, and RPA expressions Eqs. \eqref{response}-\eqref{EXX_PW} has been implemented in GPAW \cite{gpaw,mortensen,gpaw-paper}, which is a Density Functional Theory code based on the projector augmented wave (PAW) method \cite{blochl}. We refer to Ref. [\onlinecite{jun2}] for details on the PAW implementation of the response function.

\section{Results}\label{results}

\subsection{Graphene on metal surfaces}
\begin{figure}[tb]
\begin{center}
 \includegraphics[scale=0.20]{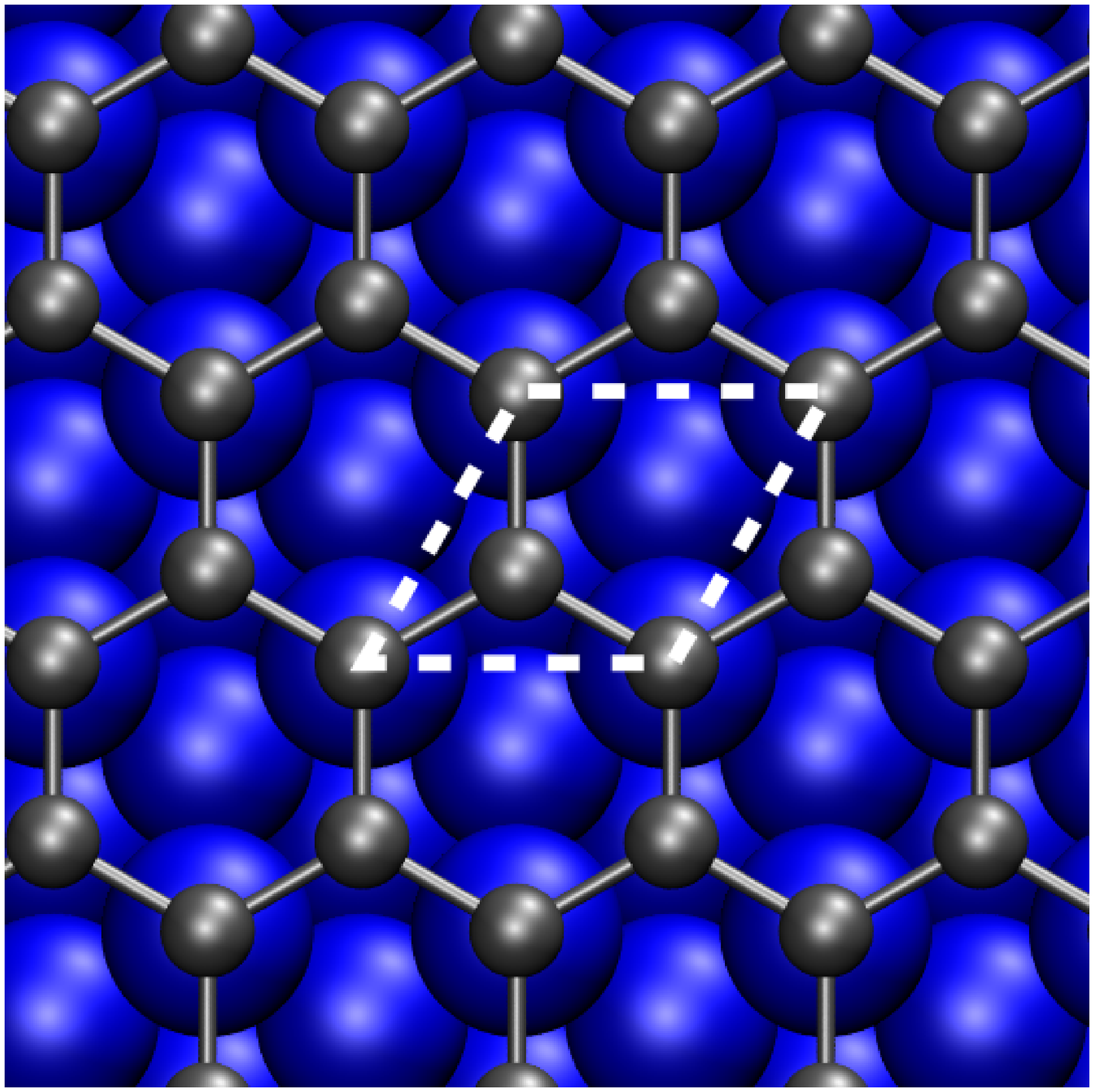}
 \hspace{0.2 in}
 \includegraphics[scale=0.21]{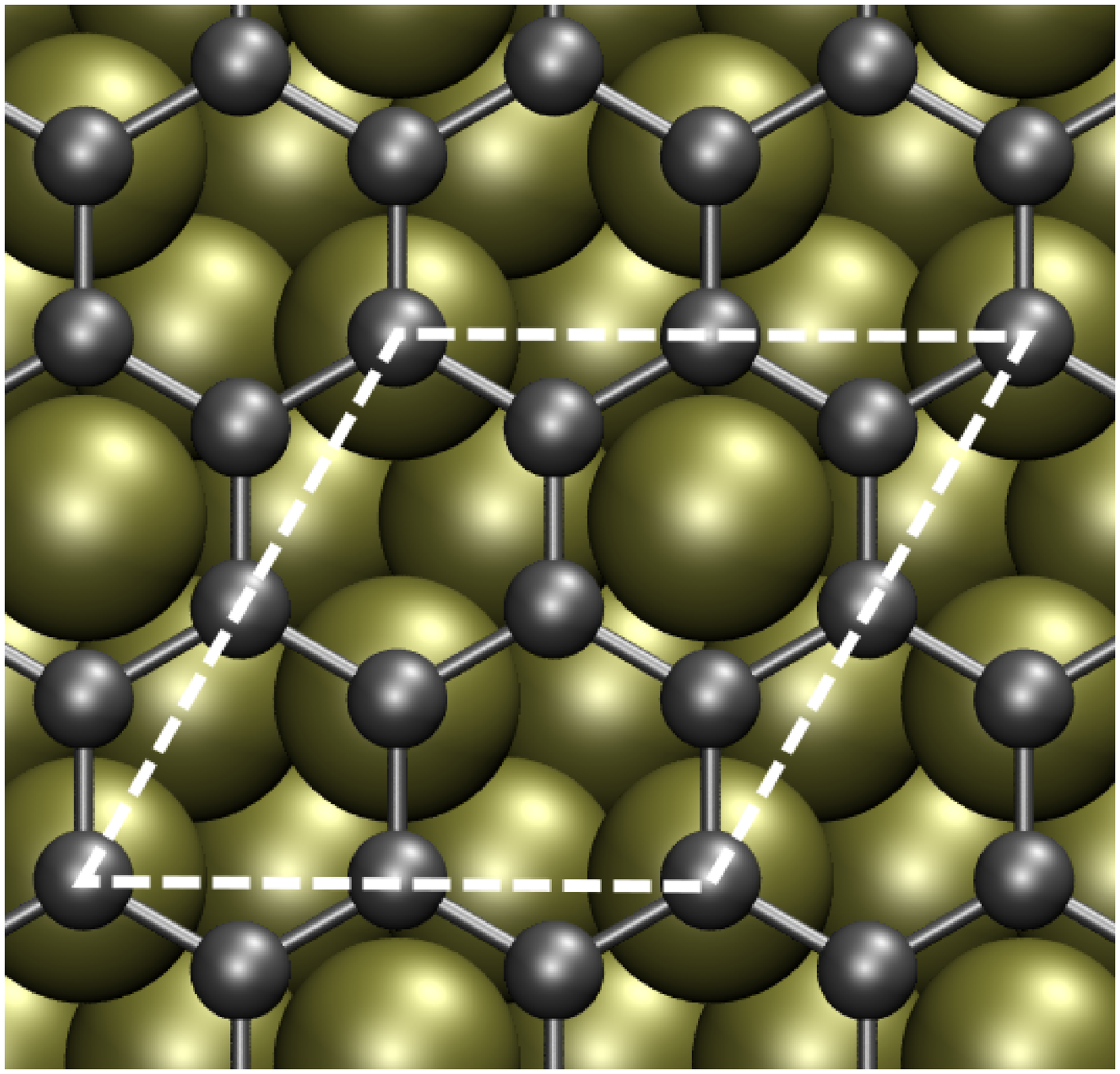}
 \caption{Left: Minimal unit cell used for Co(0001), Cu(111) and Ni(111). Right: $\sqrt{3}\times\sqrt{3}$ unit cell used for Pd(111), Pt(111), Au(111), Ag(111), and Al(111). Note that the orientation of the metal surfaces with respect to the graphene sheet differ by 30$^\circ$ in the two cases.}
\label{graph_cell}
\end{center}
\end{figure} 

The main result of the present paper is the RPA calculation of potential energy surfaces of graphene on 8 different metal surfaces. The calculation has already been carried out for Ni(111)\cite{olsen1, mittendorfer}, Cu(111) and Co(0001)\cite{olsen1} which all can be done with the minimal $1\times1$ surface unit cell. Here we extend the calculation to include Pd(111), Pt(111), Au(111), Ag(111), and Al(111), where the $\sqrt{3}\times\sqrt{3}$ surface unit cell is needed in order to obtain a periodic system which is compatible with the graphene lattice distance. In all calculations we have used the experimental metal lattice parameter, and stretched or squeezed graphene to match the unit cell. In Table \ref{tab:stretching} we present applied lattice parameters along with the energy required to stretch an isolated graphene sheet to match the lattice. For Ni, Co, and Cu we have $a=d$ and for Ag, Au, Pd, Pt, and Al we have $a=2d/\sqrt{3}$ for a metal nearest neighbor distance $d$.  
\begin{table}[tb]
\begin{center}
\begin{tabular}{c|c|c|c|c|c|c|c|c}
Metal: & Ni & Co & Cu & Pd & Pt & Au & Ag & Al \\
\hline
\textit{a} ({\AA}) & 2.49 & 2.51 & 2.56 & 2.38 & 2.40 & 2.50 & 2.51 & 2.48\\
$\Delta$E (meV)    &    6 &  21  &   92 & 91   &   52 & 13 & 21 & 2
\end{tabular}
\end{center}
\caption{Value of the graphene lattice parameter \textit{a} when matched to the various metal surfaces with experimental lattice parameters. We also display the energy per C atom $\Delta$E (calculated with the PBE functional) needed to stretch an isolated graphene sheet to the metallic lattice parameter. The experimental lattice parameter of isolated graphene is 2.46 {\AA}.}
\label{tab:stretching}
\end{table}
In the case of Cu, Pd and Pt the deviation from the applied unit cell is particularly bad and it is expected that a more complicated Moire pattern is needed in order for the graphene and metal surface to come in registry. Nevertheless, it is interesting to compare the performance of different functionals even though these structures are not observed experimentally.

For all calculations except the RPA correlation energy, we used a plane wave cutoff of $600\;eV$. For the RPA correlation energy we used the orbitals and eigenvalues obtained with $600\;eV$ cutoff and evaluated the response function at a cutoff of $200\;eV$ for the small systems and $150\;eV$ for the large systems. The number of bands included in the response function were set equal to the number of plane waves defined by the cutoff energy. The metal surface was simulated using four atomic layers and the repeated images were separated by 20 {\AA} of vacuum. For the Ni(111) and Co(0001) slabs the calculations were spin-polarized. A $16\times16$ gamma-centered k-point mesh was used for Ni and Co whereas an $12\times12$ grid was used for Cu and $6\times6$ grids were used for the large systems. We return to the issue of k-point and cutoff convergence in appendix \ref{Ni_convergence}. 

\begin{figure*}[tb]
\begin{center}
 \includegraphics[scale=0.32]{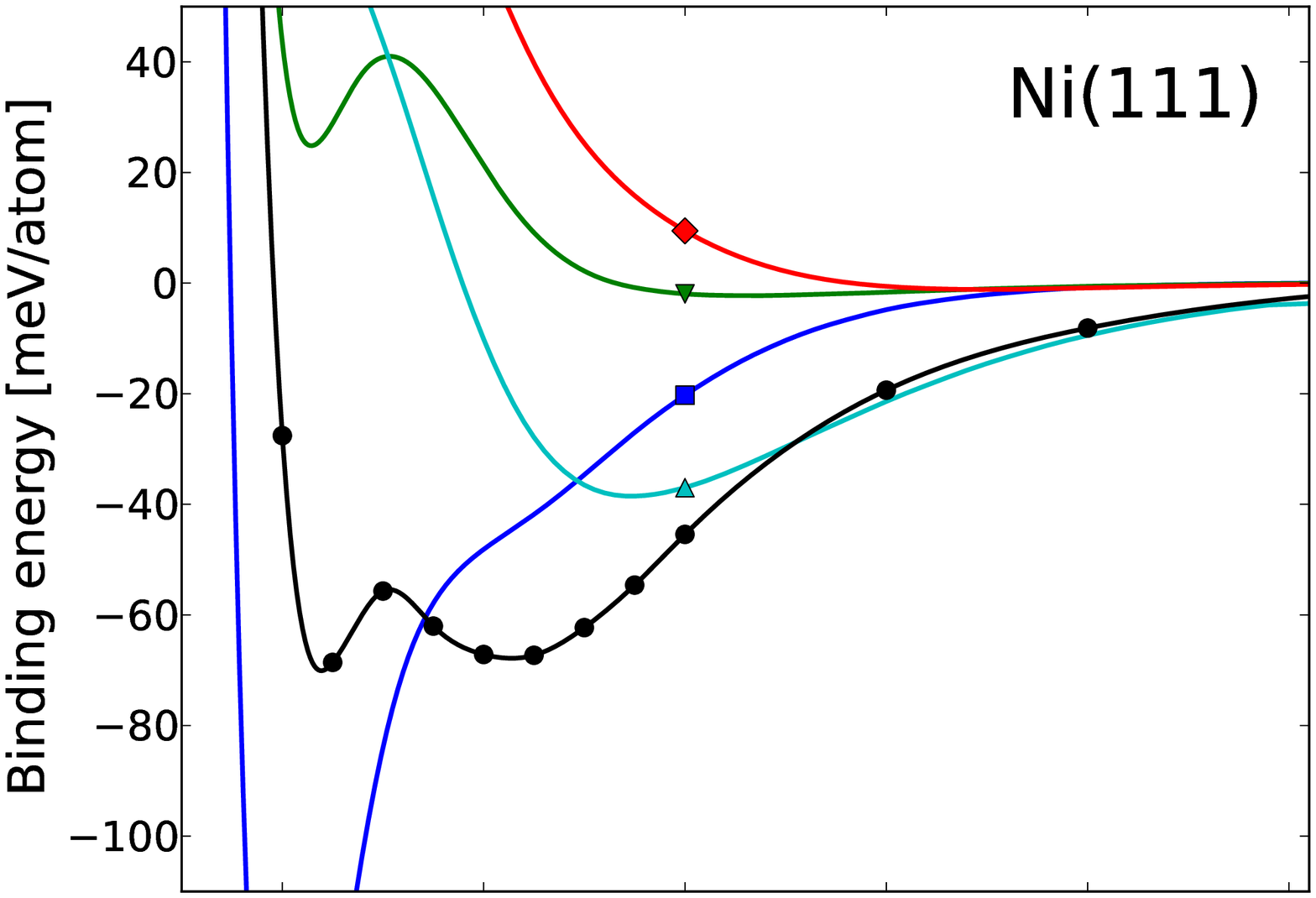}
 \includegraphics[scale=0.32]{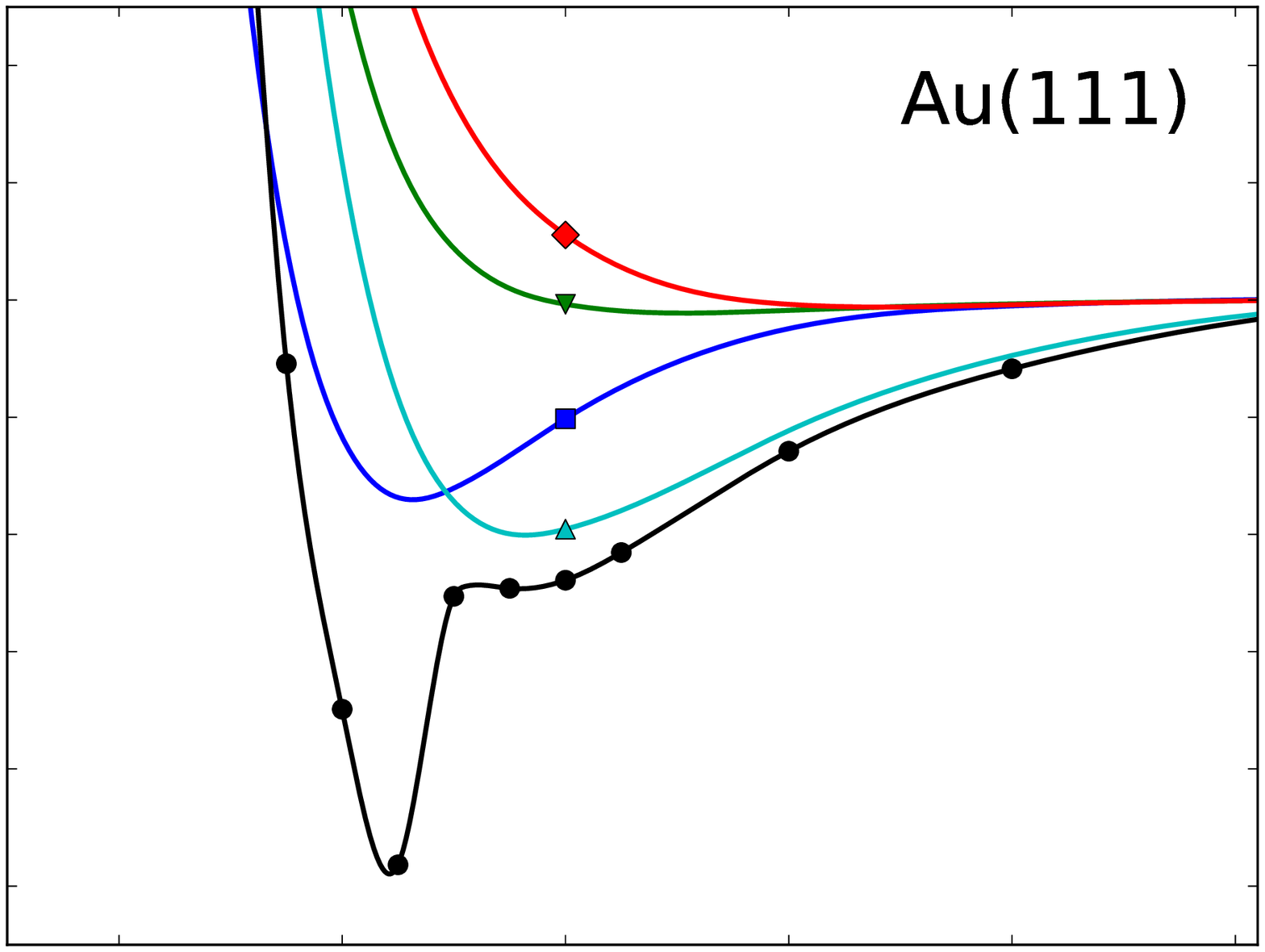}
 \includegraphics[scale=0.32]{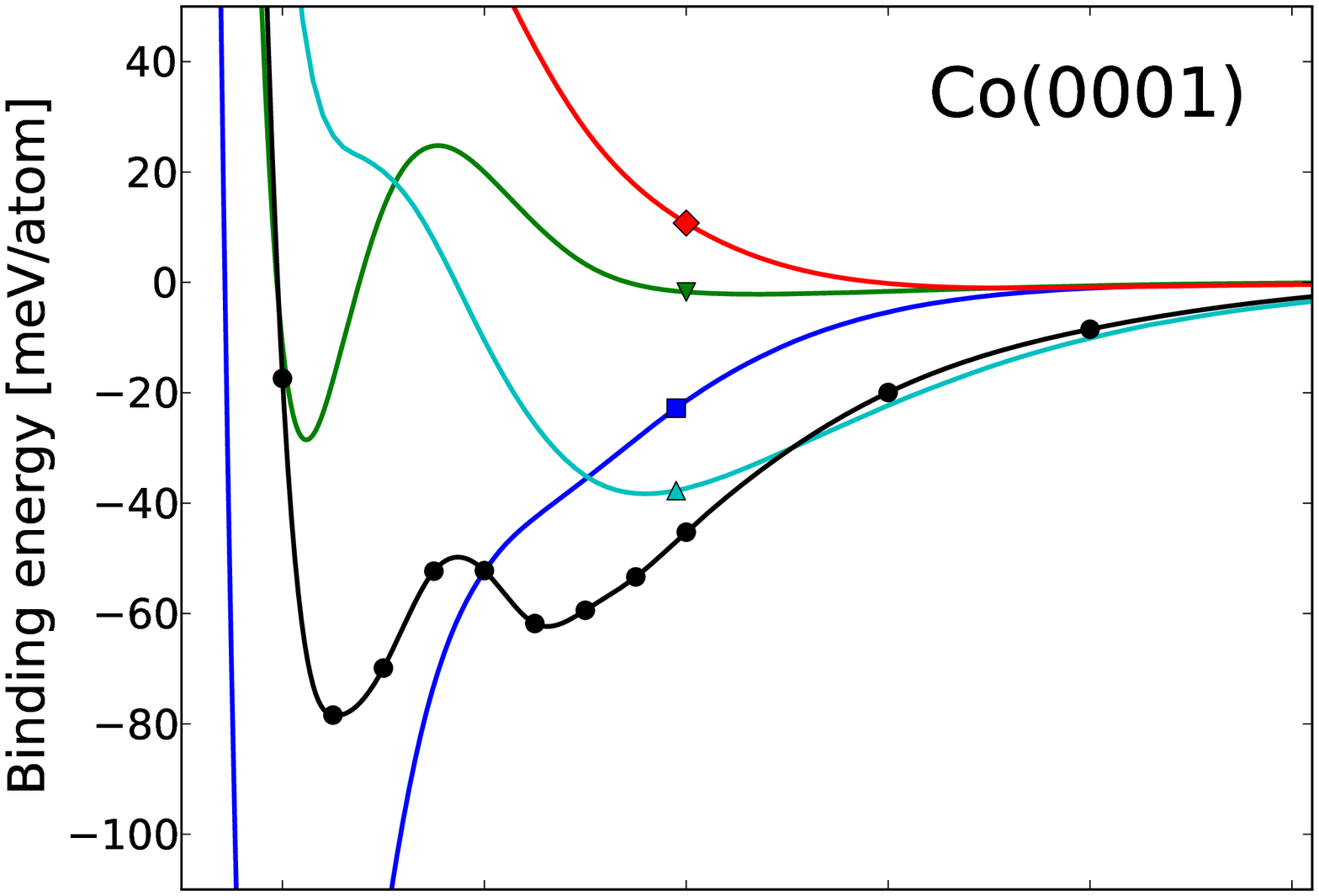}
 \includegraphics[scale=0.32]{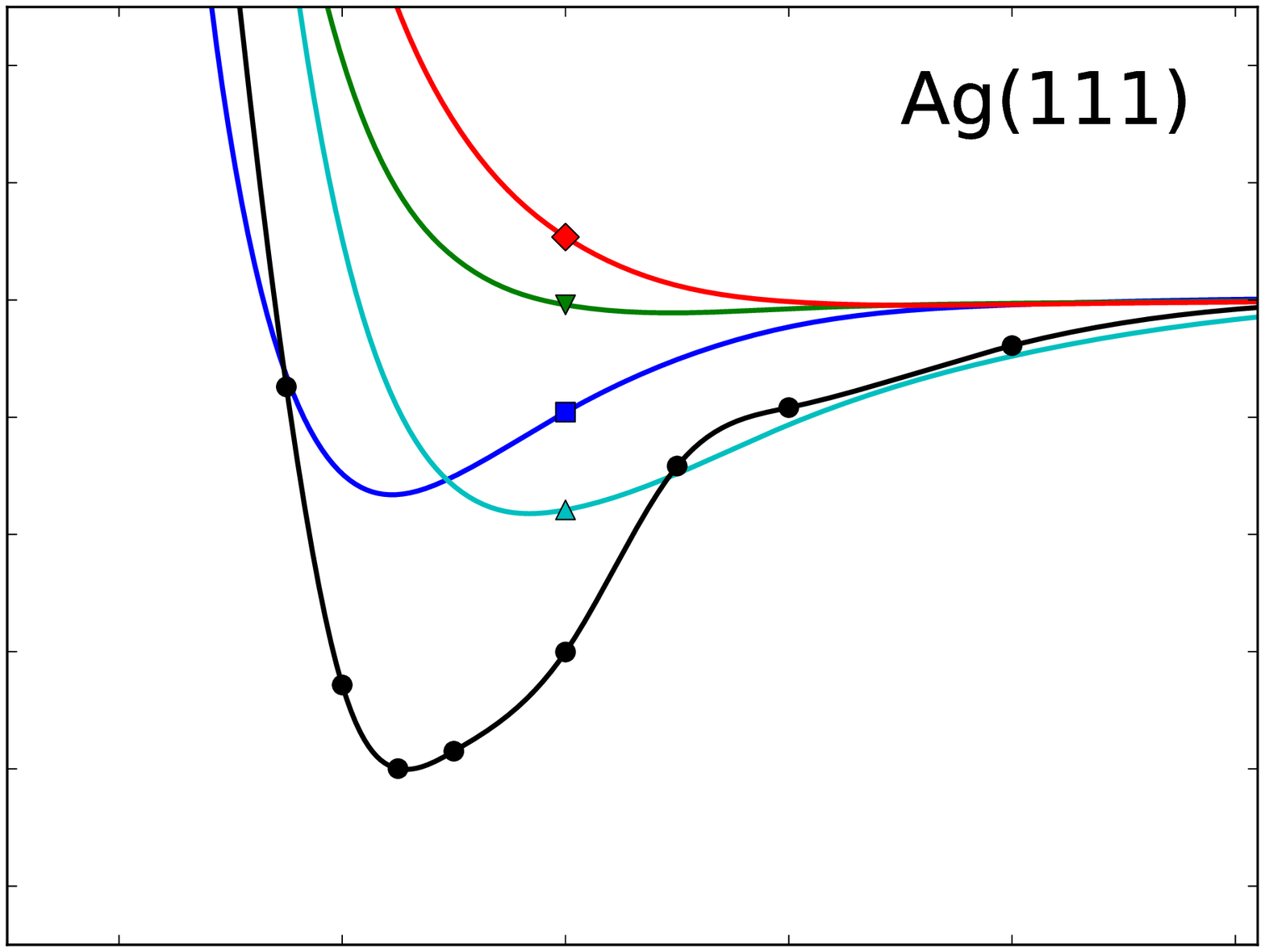}
 \includegraphics[scale=0.32]{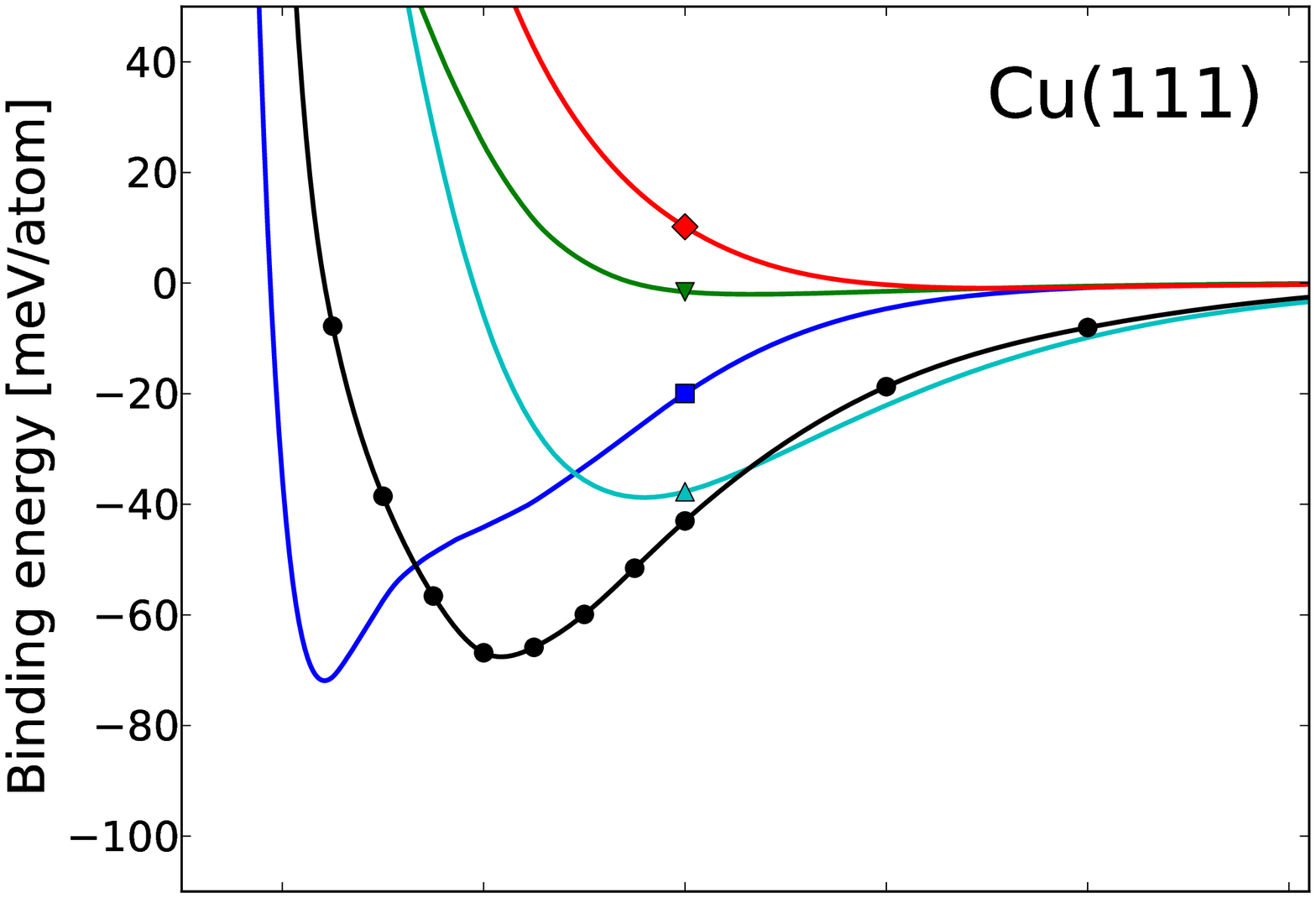}
 \includegraphics[scale=0.32]{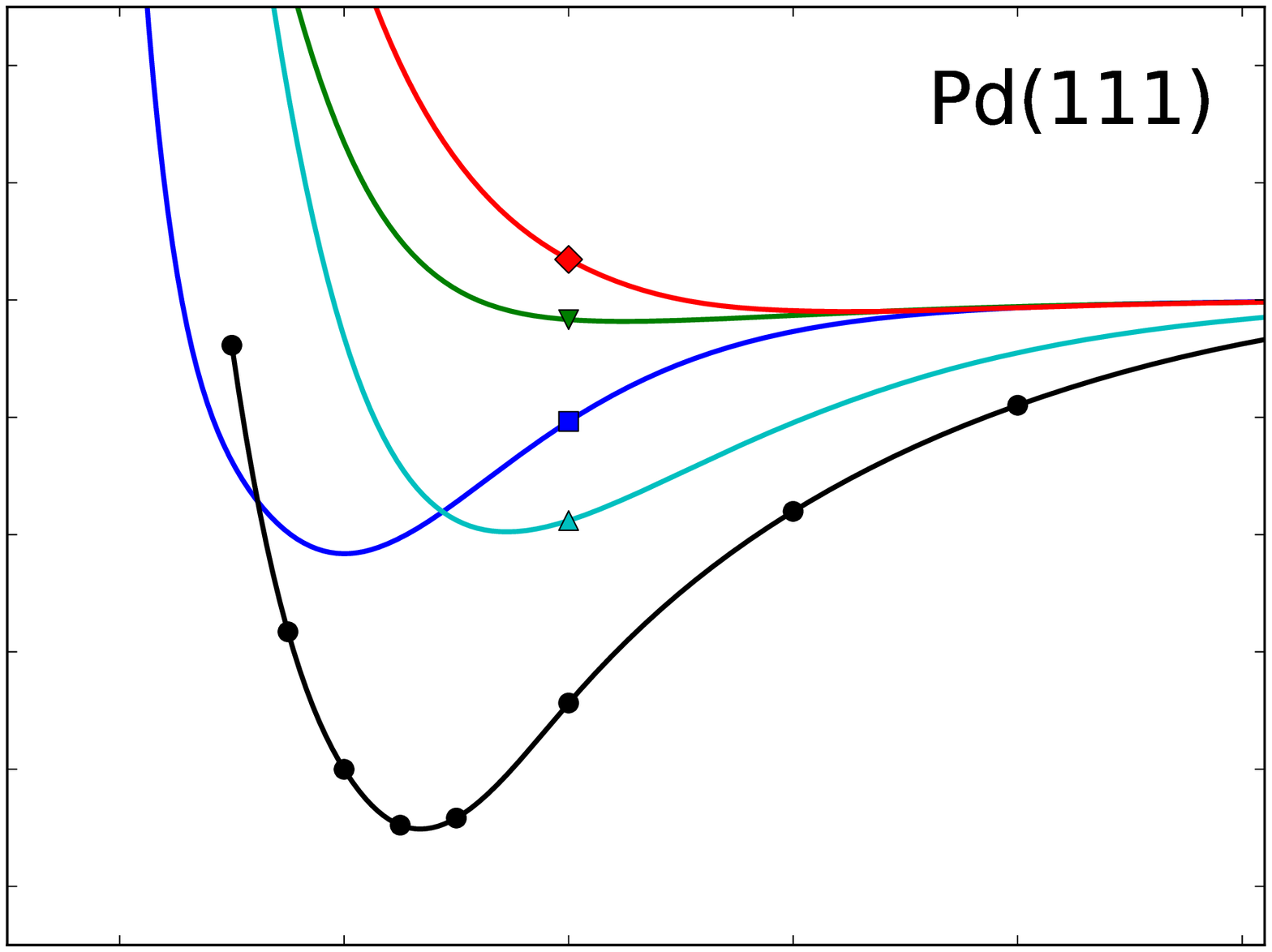}
 \includegraphics[scale=0.32]{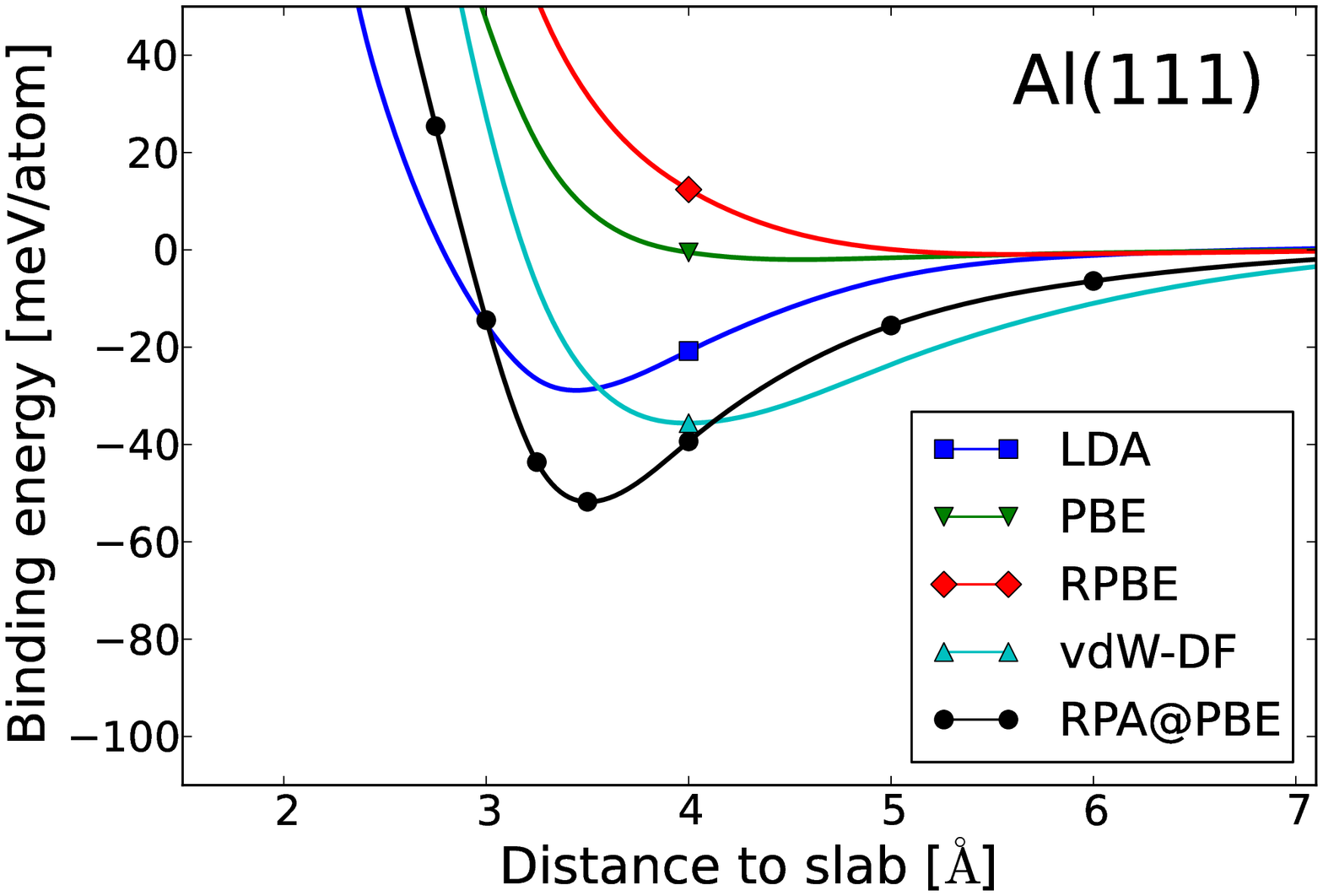}
 \includegraphics[scale=0.32]{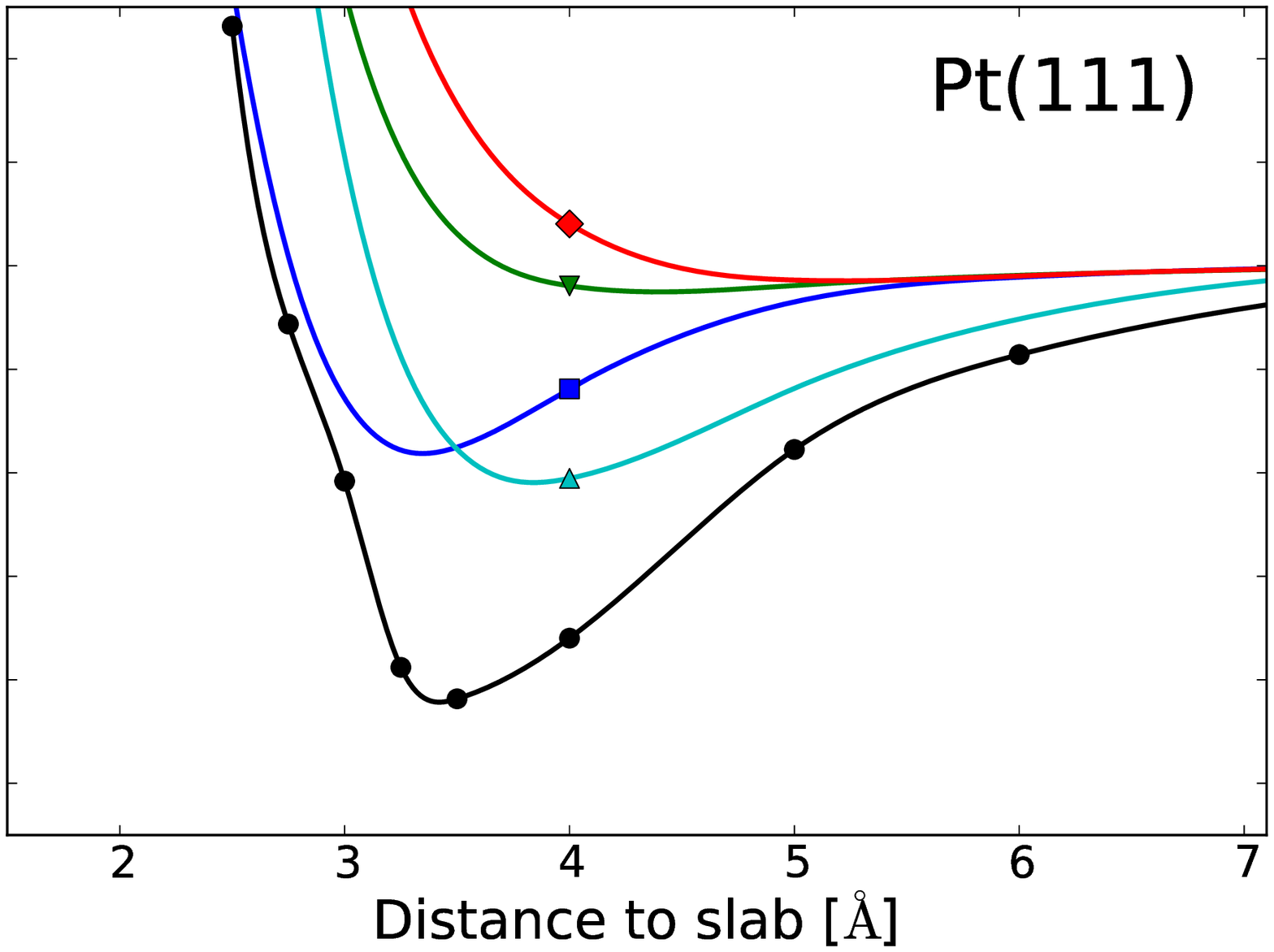}
 \caption{Potential energy curves of graphene on metal surfaces}
\label{graph_metal}
\end{center}
\end{figure*} 
In Fig. \ref{graph_metal} we show the potential energy curves of graphene on the Ni(111), Co(0001), Cu(111), Au(111), Ag(111), Pt(111), Pd(111), and Al(111) calculated with LDA, PBE, RPBE, a van der Waals functional\cite{dion} (vdW-DF), and RPA. The binding energies and equilibrium distances are summarizes in Tabs. \ref{tab:energies} and \ref{tab:distances}. LDA predicts strong binding ($\sim70-260$ meV) and small binding distance (2.0-2.21 {\AA}) for Ni(111), Co(0001) and Cu(111) and weak binding at $\sim3.3$ {\AA} for the rest of the metals. It should be noted that if the LDA optimized lattice constant is used for Cu the binding is weaker and similar to the LDA curve of Pd(111).\cite{kelly} 
PBE predicts a very weak bond at $\sim4.4$ {\AA} except in the case of Co(0001) where a minimum close to the surface ($\sim2.0$ {\AA}) appears. This feature is also observed for Ni(111) where a local minimum appears close to the surface, but in this case it is unstable with respect to the desorbed graphene. The RPBE functional predicts very weak binding far from the surface ($\sim4.4$ {\AA}) for all the systems. The vdW-DF also gives very similar results for all the metals with an equilibrium distance of $\sim3.75$ {\AA} and a binding energy of $\sim40$ meV. For Ni(111) and Co(0001) RPA produces two distinctive minima at $\sim2.2$ {\AA} and $\sim3.25$ {\AA} and in both cases the global minimum is the one close to the surface. For the rest of the systems RPA predicts an equilibrium distance at $\sim3.3$ {\AA}, but with much larger binding energies than any of the other functionals.

It has previously been shown that the electronic structure of graphene adsorbed on Ni(111) and Co(0001) is severely modified at the minimum close to the surface, whereas the graphene electronic states do not hybridize with the metallic states at $\sim3.25$ {\AA}.\cite{olsen1} Both the small binding distances and the modified electronic structure at these distances are in good agreement with experiments.\cite{eom,varykhalov} For the remaining systems a direct comparison with experiments is not possible, since extended Moire patterns are observed due to a mismatch of lattice parameters. Nevertheless, we can use these systems as a test set for comparing the performance of different functionals. RPA and vdW-DF is the only non-local functionals considered and they both capture the slowly decaying long distance tail originating from dispersive interactions. However, at intermediate distances where both dispersive and covalent interactions are important the two functionals deviate significantly. The vdW-DF gives very similar results for all systems and does not seem to capture the differences in surface electronic structure when the surface is approached. In fact, changing the local part of the vdW-DF has been shown to give rise to qualitatively different energy curves,\cite{mittendorfer, wellendorff}. Thus choosing an accurate van der Waals functional for these kind of systems only becomes possible when the accurate result is already known. While such an approach cannot really be regarded as "first principles calculations", it could be very useful for comparing certain classes of systems, once a single calculation has been benchmarked against a reliable result.\cite{mittendorfer, wellendorff} On the other hand, RPA constitutes a unique functional that do not involve arbitrary choices for exchange and (local) correlation. Since our approach to RPA is not self-consistent, there may be a dependence on the choice of orbitals and eigenvalues used to evaluate the response function. However, in our experience these differences are rather small and does not give rise to qualitative differences.

We note that one would expect the RPA energy curves at long distances to be well described by vdW-DF functionals since the local contributions to exchange and correlation then vanishes. This seems to be case for all the metals except Pd(111) and Pt(111). It is interesting that for these two metals the pure HF energy curves produce weak minima at 5 {\AA} with binding energies of 5 and 8 meV respectively. For all other surfaces the HF energy curves are purely repulsive in this region. The exchange functional used in the vdW-DF considered here is the revPBE.\cite{revpbe} This is very similar to the RPBE functional, which gives completely similar structure at large distances from the surface. It is thus very likely that the deviations between RPA and vdW-DF at a distance of $\sim 5-6$ {\AA} from the surface is due to small local exchange-correlation effects, which are not well described by the present vdW-DF. The fact that (semi) local exchange-correlation effects are important at distances of $\sim 5$ {\AA} for the surface is also supported by recent calculations with the M06-L functional,\cite{m06l} which accurately reproduce the RPA energy curve for Ni(111).\cite{andersen}
\begin{table}[tb]
\begin{center}
\begin{tabular}{c|c|c|c|c|c|c|c|c}
Metal: & Ni & Co & Cu & Pd & Pt & Au & Ag & Al \\
\hline
$E_B^{LDA}$ (meV)    & 188 & 259 & 72 & 43 & 36 & 34 & 30 & 29\\
$E_B^{PBE}$ (meV)    &   2 &  29 &  2 &  4 &  5 &  2 &  2 & 2 \\
$E_B^{RPBE}$ (meV)   &   1 &   1 &  1 &  2 &  3 &  1 &  1 & 1 \\
$E_B^{vdW}$ (meV)    &  39 &  38 & 39 & 40 & 42 & 40 & 36 & 36 \\
$E_B^{RPA}$ (meV)    &  70 &  78 & 68 & 90 & 84 & 95 & 78 & 52 \\
\end{tabular}
\end{center}
\caption{Binding energies per C atom at the equilibrium distance to the surface calculated with different functionals.}
\label{tab:energies}
\end{table}

\begin{table}[tb]
\begin{center}
\begin{tabular}{c|c|c|c|c|c|c|c|c}
Metal: & Ni & Co & Cu & Pd & Pt & Au & Ag & Al \\
\hline
$d_B^{LDA}$ ({\AA})    & 2.00 & 2.01 & 2.21 & 3.00 & 3.35 & 3.32 & 3.22 & 3.44\\
$d_B^{PBE}$ ({\AA})    & 4.33 & 2.12 & 4.33 & 4.25 & 4.40 & 4.53 & 4.47 & 4.55 \\
$d_B^{RPBE}$ ({\AA})   & 5.48 & 5.60 & 5.54 & 5.26 & 5.22 & 5.43 & 5.57 & 5.61 \\
$d_B^{vdW}$ ({\AA})    & 3.73 & 3.80 & 3.80 & 3.73 & 3.84 & 3.82 & 3.84 & 3.99 \\
$d_B^{RPA}$ ({\AA})    & 2.19 & 2.27 & 3.09 & 3.34 & 3.42 & 3.22 & 3.31 & 3.51 \\
\end{tabular}
\end{center}
\caption{Equilibrium distances to the surface calculated with different functionals.}
\label{tab:distances}
\end{table}

\subsection{Graphite}
A very important accomplishment of the RPA method is the demonstration of a correct description of the cohesive properties of graphite.\cite{lebegue} RPA gives excellent agreement with the experimental interlayer distance and interlayer binding energy which so far only has been calculated accurately with quantum Monte Carlo methods.\cite{spanu} Here we reproduce the main results of Ref. [\onlinecite{lebegue}] in order to assess the performance of the present implementation.

\begin{figure}[tb]
\begin{center}
 \includegraphics[scale=0.35]{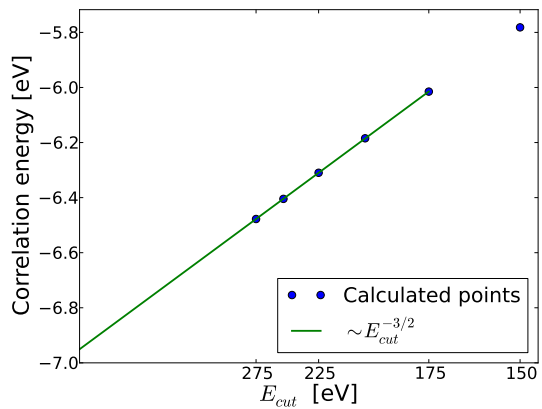}
  \caption{RPA correlation energy per atom as a function of cutoff energy at $d=3.34$ \AA. For cutoff energies above 175 eV the extrapolation scheme is rather accurate and the uncertainty on the extrapolated result is on the order of 5 meV and 1 meV for two-point extrapolation and linear regression respectively.}
\label{graphite_extra}
\end{center}
\end{figure}

For the graphite calculations we used a Gamma centered $26\times26\times8$ $k$-point grid for the DFT and HF calculations and a $14\times14\times6$ $k$-point for the RPA calculations. We have used a plane wave energy cutoff of 800 eV for the HF and DFT calculations. To obtain the RPA interlayer binding energy of graphite, one has to compare correlation energies evaluated in different unit cells corresponding to different values of the interlayer separation $d$. Therefore, one cannot use a single low energy cutoff and rely on error cancellation between energy differences as e.g. in the case of graphene on metal surfaces. Instead, for a given value of $d$ one has to obtain the converged value of the correlation energy corresponding to infinite cutoff energy. As shown empirically in Ref. [\onlinecite{harl08}], the correlation energy at high values of the cutoff scales as $\sim E_{cut}^{-3/2}$ and the converged RPA correlation energy can be obtained by fitting the function
\begin{align}\label{fit}
 E^{RPA}(E_{cut})=E^{RPA}+\frac{A}{E_{cut}^{3/2}},
\end{align}
to a sequence of cutoff energies and the associated correlation energies. In Fig. \ref{graphite_extra} we show the extrapolation at the equilibrium distance of $d=3.34$ \AA. The fit to Eq. \eqref{fit} appears rather accurate and the extrapolated correlation energy ranges from $-6.958$ eV to $-6.949$ eV when any two points between $E_{cut}=175$ eV and $E_{cut}=275$ eV are used. However, to obtain a meaningful binding energy curve, an accuracy of $\sim 1$ meV is needed and a two-point extrapolation is not sufficient. Instead, we perform linear regression on the five points: $E_{cut}\in\{175,200,225,250,275\}$ resulting in the extrapolated correlation energy $E^{RPA}=6.955\pm0.00095$ eV.

\begin{figure}[tb]
\begin{center}
 \includegraphics[scale=0.4]{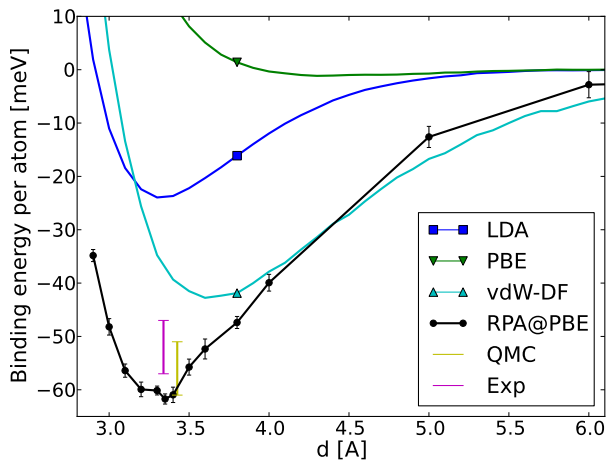}
  \caption{Potential energy surfaces for graphite obtained with RPA, LDA, PBE and a van der Waals functional. The error bars on the RPA calculations are obtained from the linear regression applied to calculate the extrapolated correlation energies (Eq. \eqref{fit})}
\label{graphite_pes}
\end{center}
\end{figure}

This extrapolation procedure is repeated for a range of different values of $d$ and the result is shown in Fig. \ref{graphite_pes} along with the results obtained with the LDA, PBE, and a van der Waals functional.\cite{dion} The result is well known: LDA predicts a fortuitous equilibrium distance, which is in good agreement with experiment and the PBE curve is purely repulsive. The van der Waals functional seems to capture part of the dispersive interactions and predicts a larger binding energy than the semi-local functionals, however, at the wrong equilibrium distance. The RPA method gives a binding energy, which is in very good agreement with experiments and quantum Monte Carlo simulations\cite{spanu} and the correct equilibrium distance at $d=3.34$ \AA. Note that we obtain a slightly larger RPA binding energy per C atom ($62$ meV) than Lebegue et al.\cite{lebegue} ($48$ meV). The reason for this could be related to the use of different PAW setups for C. It is also possible that our $k$-point sampling is not completely converged, since if we use the same $k$-point sampling ($14\times14\times6$) for both Hartree-Fock and RPA correlation we get a binding of $47$ meV per Carbon atom. However, with this $k$-point sampling the large distance tail of the Hartree-Fock PES is not converged and the total PES exhibits a spurious maximum at $d\sim5$ \AA.

\subsection{Cohesive energies and lattice constants of solids}
In Refs. [\onlinecite{harl09, harl10}] the RPA method was demonstrated to yield bulk lattice constants in very good agreement with experiment and cohesive energies somewhat worse than the PBE functional. The absolute RPA correlation energy is typically overestimated by 25-50 \%\cite{ren_review, eshuis} for atoms and molecules and slightly less for solids, but RPA energy differences are reproduced accurately when systems with similar electronic structure are compared. Lattice constants are determined by comparing very similar electronic systems and are therefore well reproduced by RPA calculations. In constrast, the computation of cohesive energies requires comparison of atoms in the solid phase with the isolated atoms, which have a completely different electronic structure and RPA performs poorly in this case.

In Tab. \ref{tab:cohesive} we display the cohesive energies of a selection of solids calculated with PBE, EXX, and RPA. In all calculations we used a gamma-centered $k$-point sampling of $12\times12\times12$ for the solid. For the calculation of the isolated atoms, the periodic images were separated by $8$ {\AA} except Na, where a separation of $10$ {\AA} was used. The RPA energy differences were calculated at different cutoff energies and extrapolated to infinity. A two-point extrapolation using Eq. \eqref{fit} with either \{250,300\} eV or \{350,400\} eV, yielded results differing by $\sim2$ meV (Si, Ge, Na) to $50$ meV (Pd, Cu). We find good agreement with the results of Ref. [\onlinecite{harl10}] with a deviation of $0.01-0.1$ eV. It should be noted that the results of Ref. [\onlinecite{harl10}] were calculated at optimized lattice constants whereas the present results are at the experimental lattice constants. This is part of the reason why our calculated cohesive energies are generally smaller than those of Ref. [\onlinecite{harl10}] and the largest deviation is seen for EXX applied to metals where a large difference from experimental lattice constants is observed. We should also remark that the PAW setups used in the present work have not been optimized for RPA calculations as in Ref. [\onlinecite{harl10}].
\begin{table}[tb]
\begin{center}
\begin{tabular}{c|c|c|c|c}
                 & PBE & EXX & RPA & Expt. \\
 & $a_{exp}\,\,\,$  $a_{opt}$  & $a_{exp}\,\,\,$  $a_{opt}$  & $a_{exp}\,\,\,$  $a_{opt}$   & \\
\hline
C   & 7.73 (7.72) & 5.16 (5.18) & 6.99 (7.00) & 7.55 \\
Si  & 4.55 (4.55) & 2.82 (2.82) & 4.32 (4.39) & 4.68 \\
SiC & 6.38 (6.40) & 4.32 (4.36) & 5.96 (6.04) & 6.48 \\
Ge  & 3.72 (3.71) & 2.05 (1.95) & 3.56 (3.59) & 3.92 \\
MgO & 4.97 (4.98) & 3.35 (3.47) & 4.85 (4.91) & 5.20 \\
Na  & 1.08 (1.08) & 0.20 (0.23) & 0.98 (1.00) & 1.12 \\
Pd  & 3.68 (3.74) &-1.44 (-1.26)& 3.51 (3.41) & 3.94 \\ 
Rh  & 5.61 (5.74) &-3.01 (-2.88)& 5.10 (5.05) & 5.78 \\ 
Cu  & 3.40 (3.48) &-0.23 (0.03) & 3.20 (3.36) & 3.52 \\
\hline
MAE & 0.16 (0.13) & 3.22 (3.14) & 0.41 (0.38) &
\end{tabular}
\end{center}
\caption{Cohesive energies of solids evaluated at the experimental lattice constant corrected for zero-point anharmonic effects. Numbers in brackets are at optimized lattice constant and are taken from Ref. [\onlinecite{harl10}]. Experimental cohesive energies are corrected for zero point energy. All numbers are in eV.}
\label{tab:cohesive}
\end{table}

In Tab. \ref{tab:lattice} we show the calculated lattice constants of C, Si, Na, and Pd. Again we find good agreement with the results of Ref. [\onlinecite{harl10}]. The results were obtained by calculating $7-9$ energy points in the vicinity of the experimental lattice constant and fitting a third order inverse polynomial to the energy-volume curve.\cite{eos1,eos2,eos3} The RPA results were obtained by a two-point extrapolation with \{250,300\} eV using Eq. \eqref{fit}. In principle, this approach also gives the bulk modulus as the curvature at the minimum. However, for the RPA calculations, the present extrapolation scheme is not accurate enough for this purpose. Linear regression involving more cutoff points would be needed in order to produce a reliably RPA bulk modulus.
\begin{table}[tb]
\begin{center}
\begin{tabular}{c|c|c|c|c}
                 & PBE & EXX & RPA & Expt. \\
\hline
C   & 3.57 (3.57) & 3.55 (3.54) & 3.57 (3.57) & 3.55 \\
Si  & 5.48 (5.47) & 5.49 (5.48) & 5.45 (5.43) & 5.42 \\
Na  & 4.20 (4.20) & 4.47 (4.49) & 4.29 (4.18) & 4.21 \\
Pd  & 3.95 (3.94) & 4.03 (4.00) & 3.90 (3.90) & 3.88
\end{tabular}
\end{center}
\caption{Optimal lattice constants of a few solids. Experimental values are corrected for zero-point anharmonic effects. All numbers are in \AA. Numbers in brackets are taken from Ref. [\onlinecite{harl10}].}
\label{tab:lattice}
\end{table}

\subsection{Dissociation of molecules}
The calculation of molecular atomization energies is not well suited for a plane wave implementation, since the number of plane waves included at a given energy cutoff, scales as the cube of the super cell size. Therefore, the dimension of the response function $\chi_{\mathbf{G}\mathbf{G}'}$ quickly becomes prohibitly large when the super cell is increased and it becomes very difficult to compute RPA correlation energies is a plane wave basis. Nevertheless, we can calculate RPA atomization energies for \textit{small} molecules and compare our implementation with codes using atomic basis sets.

To obtain the RPA correlation part of atomization energies, we use Eq. \eqref{fit} to extrapolate calculations performed at $E_{cut}\in\{150,200,250,300,350,400\}$ eV. While the absolute RPA correlation energies are hard to converge, extrapolated energy differences are converged when points in the range $E_{cut}\in\{300,350,400\}$ are used. Thus, if the same unit cell is used for the calculation of the molecular correlation energy and the atomic correlation energies, the energy difference can be obtained by a two-point extrapolation using either $E_{cut}\in\{350,400\}$ eV or $E_{cut}\in\{300,350\}$ eV. The extrapolated energy differences differ by at most $20$ meV depending on which two cutoff energies are used. In contrast, the individual correlation energies of atoms and molecules are much harder to converge and linear regression is needed in order to get a reliable extrapolated result. In general, larger unit cells tend to improve the accuracy of the extrapolation, since the larger number of plane waves results in a smoother cutoff dependence. In Appendix \ref{CO_convergence} we show various convergence test for the correlation energy of the CO molecule.

\subsubsection{Atomization energies}
We have computed the atomization energies of 12 small molecules and compared with the results obtained by Furche\cite{furche,furche_voorhis} using an atomic basis set approach. The calculations were performed on experimental geometries and the experimental atomization energy has been corrected for zero-point vibrational energies. For P and Cl the calculations were performed in a supercell where the nearest neighbor atoms of periodic images were separated by 8 {\AA}. For the rest of the elements a separation of 6 {\AA} was sufficient. The results are shown in Tab. \ref{tab:atomization} and we observe a close agreement with the results of Furche. 
\begin{table}[t]
\begin{center}
\begin{tabular}{c|c|c|c|c|c}
      & LDA  & PBE & RPA - This work & RPA - Ref. [\onlinecite{furche_voorhis}] & Exp.\\
	\hline
H$_2$     & 113 & 105 & 109 & 109 & 109 \\
N$_2$     & 268 & 244 & 224 & 223 & 229 \\
O$_2$     & 174 & 144 & 112 & 113 & 121 \\
CO        & 299 & 269 & 243 & 244 & 259 \\
F$_2$     &  78 &  53 &  30 &  31 &  39 \\
HF        & 161 & 142 & 131 & 133 & 141 \\
H$_2$O    & 266 & 234 & 222 & 224 & 232 \\
C$_2$H$_2$& 460 & 415 & 383 & 381 & 405 \\
CH$_4$    & 462 & 420 & 404 & 405 & 419 \\
NH$_3$    & 337 & 302 & 290 & 290 & 297 \\
Cl$_2$    &  81\footnote{We were not able to converge the LDA energy of the isolated Cl atom and the non-selfconsistent LDA energy evaluated at the PBE density was used here.} &  65 &  49 &  50 &  58 \\
P$_2$     & 143 & 121 & 116 & 116 & 117 \\
\hline
MAE       &  36 &   8 &   9 &   9 &
\end{tabular}
\end{center}
\caption{Atomization energies of small molecules. The results from Ref. [\onlinecite{furche_voorhis}] were performed with atomic orbital basis set. All RPA energies were performed with selfconsistent PBE orbitals and eigenenergies. All numbers are in kcal/mol (1 kcal/mol=43 meV). }
\label{tab:atomization}
\end{table}

The comparison with experimental values and the PBE functional is well known. RPA systematically underestimates atomization energies and performs slight worse than the PBE functional, but significantly better than LDA. 

\subsubsection{Static correlation of MgO dimer}
An accurate description of the MgO dimer ground state energy, represents a challenge for any single reference \textit{ab initio} method due to the multi reference nature of the ground state.\cite{thummel, maatouk} The two lowest lying electronic states are the singlet $\text{X}^1\Sigma^+$ and triplet $\text{a}^3\Pi$ with the former being favored by 0.2 eV. Both of these states have an "open shell" ionic character with Mg donating an electron to O and are correlated with ionic diabatic dissociation limits.\cite{thummel} This results in a significant hybridization between valence and Rydberg states and the dimer in its ground states is not well characterized by a single Slater determinant. Moreover, while the adiabatic potential energy curve for the triplet naturally dissociates into the lowest lying atomic configuration $^1\text{Mg}+^3\text{O}$, the singlet will dissociate into $^1\text{Mg}+^1\text{O}$. Here we examine how the the atomization energy of the lowest singlet and triplet states are described by Hartree-Fock and RPA. Thus we calculate the atomization energies $E_a=E_\text{Mg}+E_\text{O}-E_{t/s}$, where $E_{t/s}$ are the ground state energies of the triplet/singlet, $E_\text{Mg}$ is the energy of a single Mg atom in its singlet state, and $E_\text{O}$ is the energy of a single O atom in the triplet state. 

In Table \ref{tab:dimer} we show the calculated atomization energies of MgO using the PBE functional, Hartree-Fock, and RPA and compare with high-level correlated methods.\cite{maatouk} The calculations were performed with fixed equilibrium geometries taken from Ref. [\onlinecite{maatouk}]. PBE overestimates the atomization energies slightly, but predicts the correct order of adiabatic states. In contrast, HF is not able to capture the static correlation originating from the ionic configuration of the dimer and predicts the singlet to be unstable. Remarkably, RPA produces correlation energies for the singlet and triplet, which differ by $\sim 3.0$ eV, but they correct the HF energies just right, such that the order of adiabatic states is restored. The total RPA atomization energies slightly underestimate the exact atomization energies, which is in line with the trend previously observed for small molecules.
\begin{table}[tb]
\begin{center}
\begin{tabular}{c|c|c|c|c|c}
      & PBE  & PBE$_\text{X}$ & HF@PBE & RPA@PBE & Ref. [\onlinecite{maatouk}]\\
	\hline
$\text{X}^1\Sigma^+$ & 2.86 & 1.86 & -1.85 & 2.48 & 2.68 \\
$\text{a}^3\Pi$      & 2.63 & 2.06 & 0.92 & 2.25 & 2.48
\end{tabular}
\end{center}
\caption{Atomization energies of the lowest singlet and triplet states of the MgO dimer. All numbers are in eV.}
\label{tab:dimer}
\end{table}

The reason for very different contributions from HF and RPA correlation, despite similar total energies, is most likely related to the second terms of Eqs. \eqref{E_x} and \eqref{E_c}. From the point of view of the adiabatic connection, the separation into exchange and correlation is somewhat arbitrary, since $\chi^{KS}$ has been added and subtracted from Eq. \eqref{E_xc}. Clearly, if $\chi^{KS}$ gives a large contribution to Eq. \eqref{E_x}, one would not expect the HF energy to be accurate. In contrast, when the RPA correlation energy is added, the contribution from  $\chi^{KS}$ cancels out and one is left with the two terms in Eq. \eqref{E_xc}. In this respect, the most surprising result in table \ref{tab:dimer} is perhaps the fact that the PBE energies are so close to the exact result. This is a manifestation of the accurate error cancellation between exchange and correlation in the PBE functional. The PBE exchange energy is very far from the exact exchange (here represented by non-selfconsistent HF), but when correlation is included the PBE and RPA approach yield similar results with the correct ordering of states. 

\subsubsection{The atomic limit of molecular dissociation}
A surprising feature of the RPA is the correct description of the atomic limit of molecular dissociation.\cite{furche,eshuis,ren_review,sanchez} Apparently, the non-perturbative nature of RPA captures the strong static correlation arising in the atomic limit, which is a remarkable property of a single reference method. For example, semi-local DFT, Hartree-Fock and coupled cluster typically yield dissociation limits, which have to high energies.\cite{ren_review} However, RPA fails dramatically in the case of H$_2^+$ dissociation, which is completely free of electronic correlation. In fact, the RPA total energy of a single H atom is $\sim-0.6$ eV and the atomic limit of H$_2$ dissociation only comes out right if taken with respect to the RPA reference of a single H atom.

\begin{figure}[tb]
\begin{center}
 \includegraphics[scale=0.35]{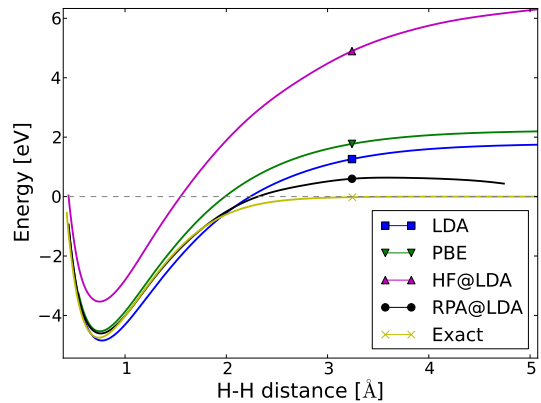}
  \caption{Dissociation curves of H$_2$. The reference energy ($E=0$) is two isolated H atoms. The exact curve is taken from Ref. [\onlinecite{wolniewicz}]}
\label{dimer_H2}
\end{center}
\end{figure}

Here we will attempt to reproduce the well known energy curves of H$_2$ dissociation using our plane wave implementation. Of course, the atomic limit of molecular dissociation is extremely difficult to reproduce with plane waves and periodic boundary conditions due to the large unit cells required. However, it is a nice test of the present implementation to see if the static correlation can be captured using plane waves and periodic boundary conditions. In the case of H$_2$, convergence is very fast with respect to cutoff and we can manage to obtain converged RPA correlation energies using extrapolation with $E_{cut}\in\{100,150\}$ eV. In Fig. \ref{dimer_H2} we show the dissociation curve of H$_2$ where we used a unit cell size of $2.5d\times2.5d\times3.5d$ with $d$ being the H-H distance. We were not able to perform calculations with unit cell sizes beyond $12\times12\times16$ {\AA}, but the trend of the RPA curve seems to agree very well with the results of Refs. [\onlinecite{eshuis,ren_review,sanchez}]. However, note the results are not completely converged with respect to unit cell size. The spurious maximum at $d=3.7$ {\AA}, also found in previous studies, is situated at $E_{max}=0.66$ eV. If we instead use unit cells of $2d\times2d\times3d$ and $3d\times3d\times3d$ we obtain $E_{max}=0.56$ eV and $E_{max}=0.72$ eV respectively. It should be noted that the energy in Fig. \ref{dimer_H2} is taken with respect to two isolated H atoms for which RPA gives $E_c^{RPA}=-0.57$ eV. On an absolute scale RPA would thus underestimate the entire energy curve by $\sim 1.1$ eV. 

\begin{figure}[tb]
\begin{center}
 \includegraphics[scale=0.35]{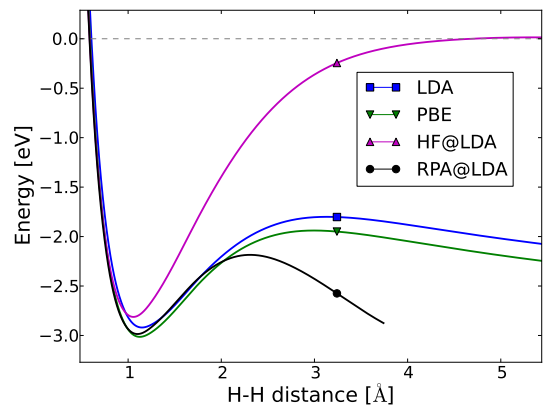}
  \caption{Dissociation curves of H$_2^+$. The reference energy ($E=0$) is an isolated H atom.}
\label{dimer_H2_plus}
\end{center}
\end{figure}

In Fig. \ref{dimer_H2_plus} we show the dissociation curve of H$_2^+$. Again, we emphasize that these molecular systems are far from the periodic systems for which the implementation was intended and we are not able to increase the H-H distance beyond 4 {\AA}. Nevertheless, our dissociation curves are in good agreement with Ref. [\onlinecite{sanchez}] and illustrates the dramatic failure of RPA for the atomic limit of open shell systems.


\section{Outlook}
In the case of adsorption of graphene on metal surfaces, RPA seem to yield results that are in better agreement with experiments than both semi-local and effective non-local vdW density functionals. This is not surprising since both covalent and dispersive interactions are important for these systems and the results seems to be in accordance with calculations for two-dimensional materials\cite{marini,lebegue,bjorkman} where RPA predicts the correct interlayer binding distance. However, it is well established that RPA does not describe covalent interactions very well and significantly underestimate the atomization energies of molecules\cite{furche} and cohesive energies of solids.\cite{harl10} One would therefore expect that the dispersive interactions (far from the surface) are very well represented, whereas the covalent interactions (close to the surfaces) are less accurate. In particular, the cases of Ni(111) and Co(0001) exhibit two minima which are very close in energy (5 and 17 meV respectively). It is highly likely that RPA underestimate the depth of the chemisorption minimum compared to the physisorption minimum and it is thus expected that the exact energy curves would have even deeper minima close to the surface. 

Since graphene on metals are being used for benchmarking new van der Waals functionals,\cite{wellendorff,andersen, mittendorfer} it is extremely important to improve the description of such systems beyond RPA. One line of development in this direction is to add a second order screened exchange term to the RPA correlation energy.\cite{gruneis} This approach exactly cancels the RPA one-electron self-correlation and improves molecular atomization energies slightly, but destroys the accurate description of static corelation in the atomic limit of molecular dissociation\cite{ren_review}. A somewhat orthogonal line is to improve the approximation for the interacting response function within TDDFT by introducing an xc kernel.\cite{lein,dobson_wang00,fuchs} The most sophisticated development in this direction is the full time-dependent EXX approach\cite{hesselmann} which is free of one-electron self-correlation, improves atomization energies compared to RPA, and reproduces the correct atomic limit of static correlation. However, this approach may easily become prohibitly heavy due to the evaluation of a frequency dependent EXX kernel and it is not clear if the method can be generalized to periodic systems. On the other hand it has been shown that the correct dynamic properties of the xc kernel is not of vital importance for total energy calculations\cite{lein} and one could simply try to use an adiabatic local xc kernel. However, as shown in Ref. [\onlinecite{furche_voorhis}], all local kernels introduce a divergence in the pair distribution function, which makes it very hard to converge correlation energies and deteriorates the accuracy of total energy calculations. We have recently shown that the divergence can be removed by a density dependent renormalization of adiabatic kernels, which defines a new class of explicit non-local adiabatic kernels.\cite{olsen2} So far this approach has been shown to significantly increase the accuracy of molecular atomization energies compared to RPA and it will very interesting to see if it performs equally well for periodic systems and graphene on metal surfaces in particular.

\begin{acknowledgments}
The authors acknowledge support from the Danish Research Council's Sapere Aude Program. The Center for
Nanostructured Graphene is sponsored by the Danish National Research Foundation.
\end{acknowledgments}

\appendix
\section{Convergence of RPA calculations}
Here we will briefly discuss a few issues regarding convergence of some of the RPA calculations presented in this paper.
 
\subsection{Graphene on Ni(111)}\label{Ni_convergence}
The computational time of RPA calculations, scales as the number of $k$-points squared, since the expression \eqref{RPA_PW} involves a sum over both $q$-points and $k$-points. Such scaling makes convergence with respect to $k$-point sampling much more cumbersome than for standard DFT calculations. In particular, \textit{ab initio} calculations of systems involving graphene may require a high $k$-point sampling to resolve the Dirac cone and RPA calculations of such systems may easily become very computationally demanding. Furthermore, It is not possible to perform an absolute convergence of the cutoff energy and extrapolation is needed in order to estimate the converged correlation energy. Since graphene on metal surfaces only bind by $\sim 50-100$ meV per C atom, the energy curves are easily destroyed by noise from the extrapolation scheme, which typically has an accuracy of $\sim 10$ meV. However, the extrapolation may be avoided when evaluating energy differences between systems of similar electronic structure, but careful convergence tests are needed to assess such behavior.

\begin{figure*}[tb]
\begin{center}
 \includegraphics[scale=0.35]{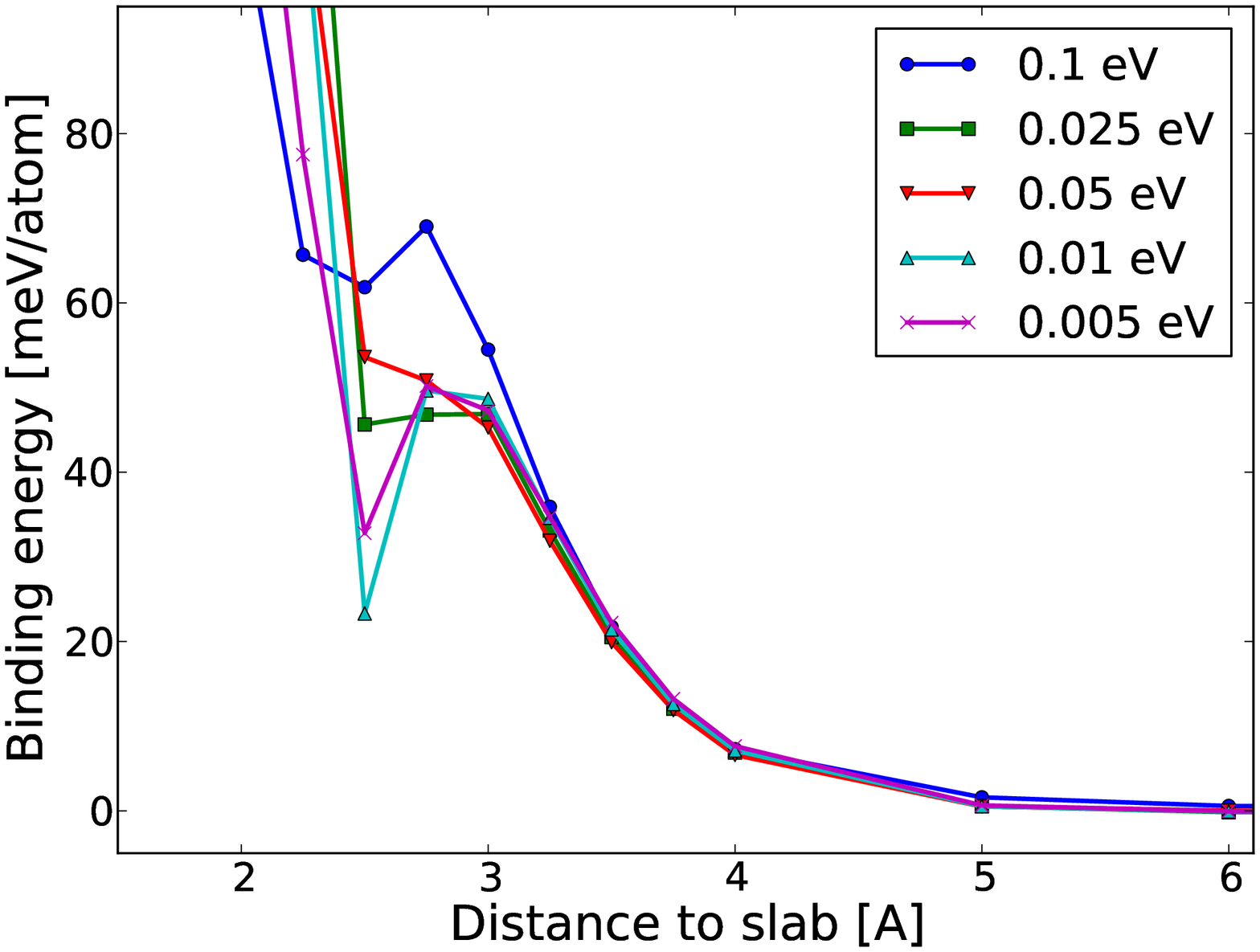}
 \includegraphics[scale=0.35]{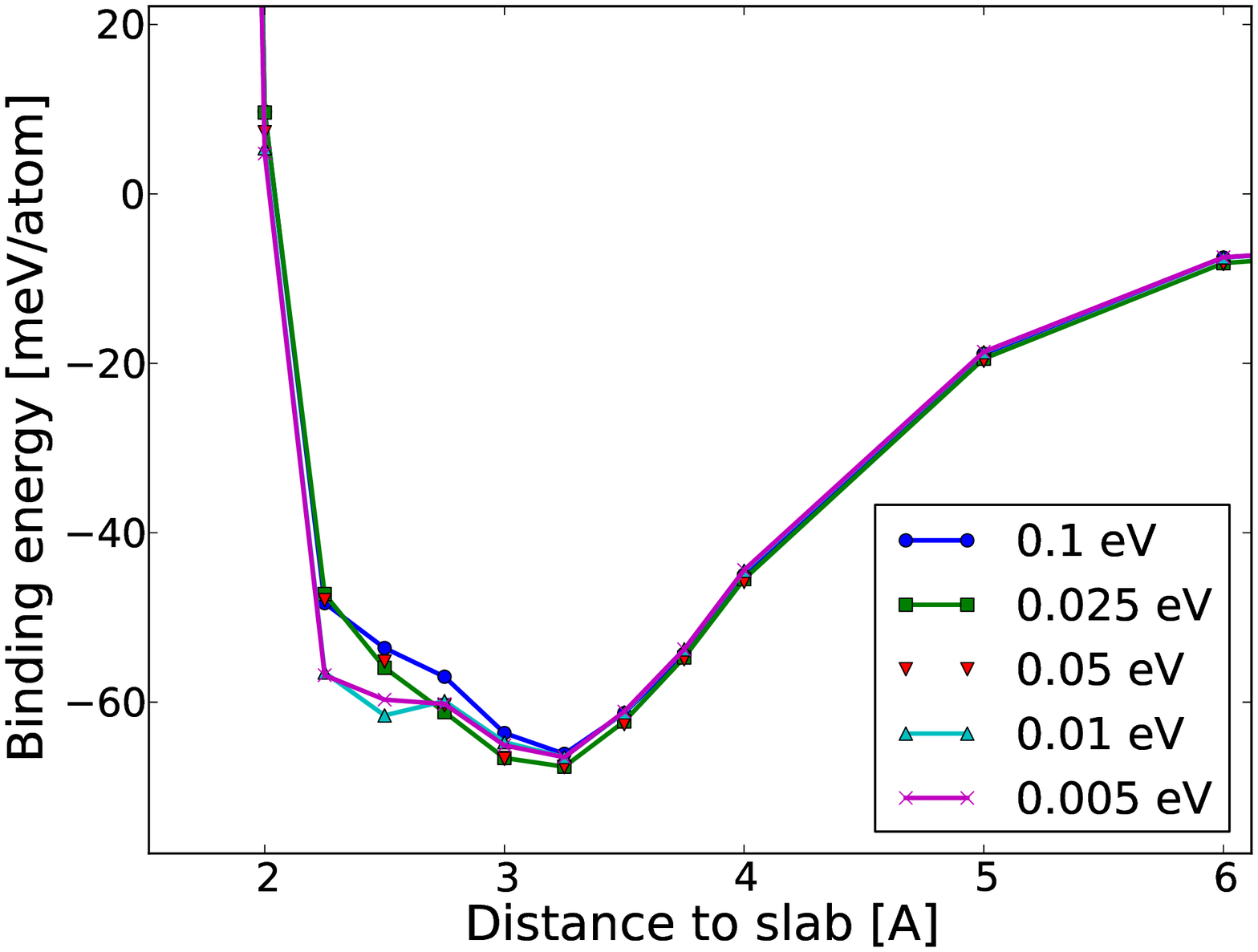}
 \includegraphics[scale=0.35]{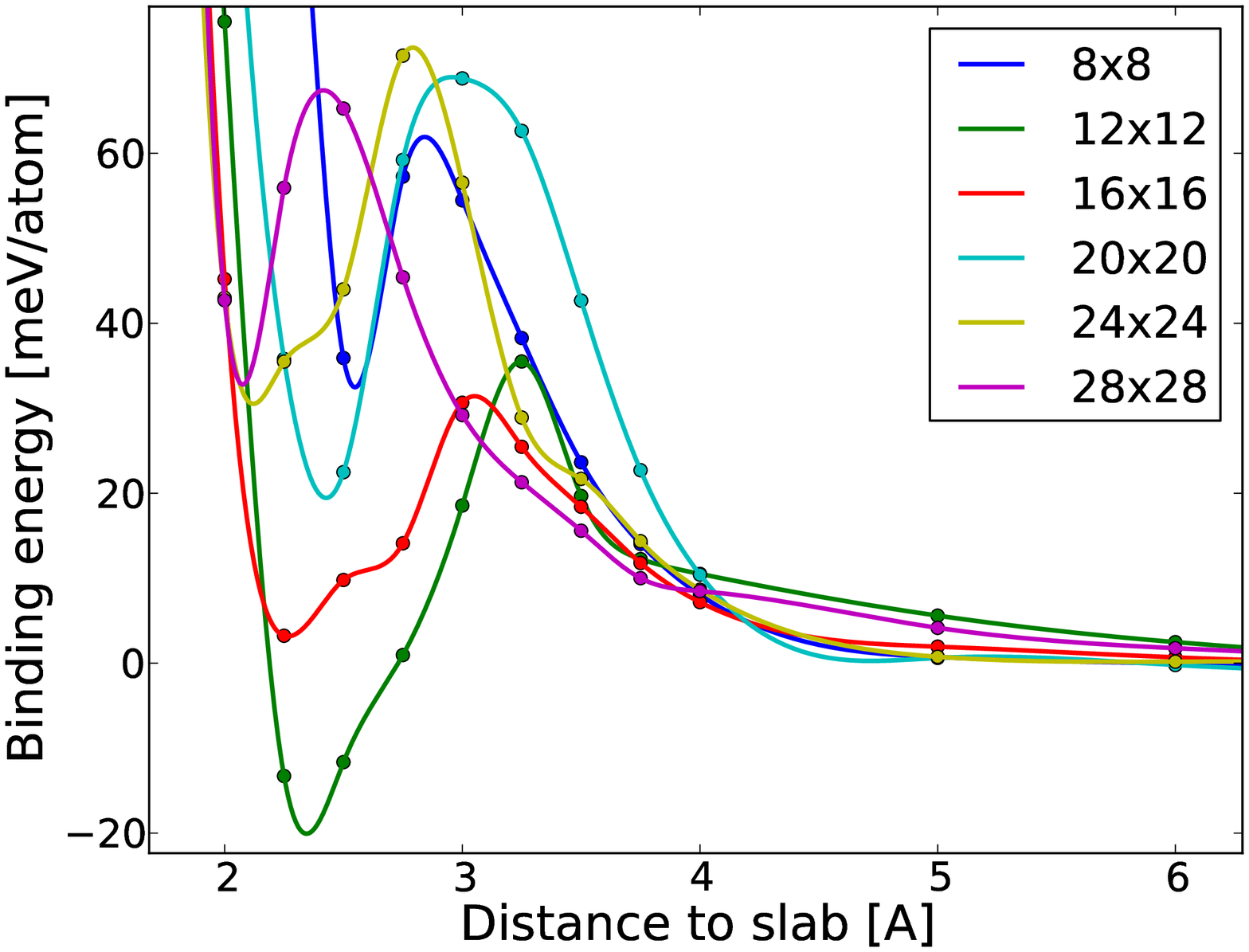}
 \includegraphics[scale=0.35]{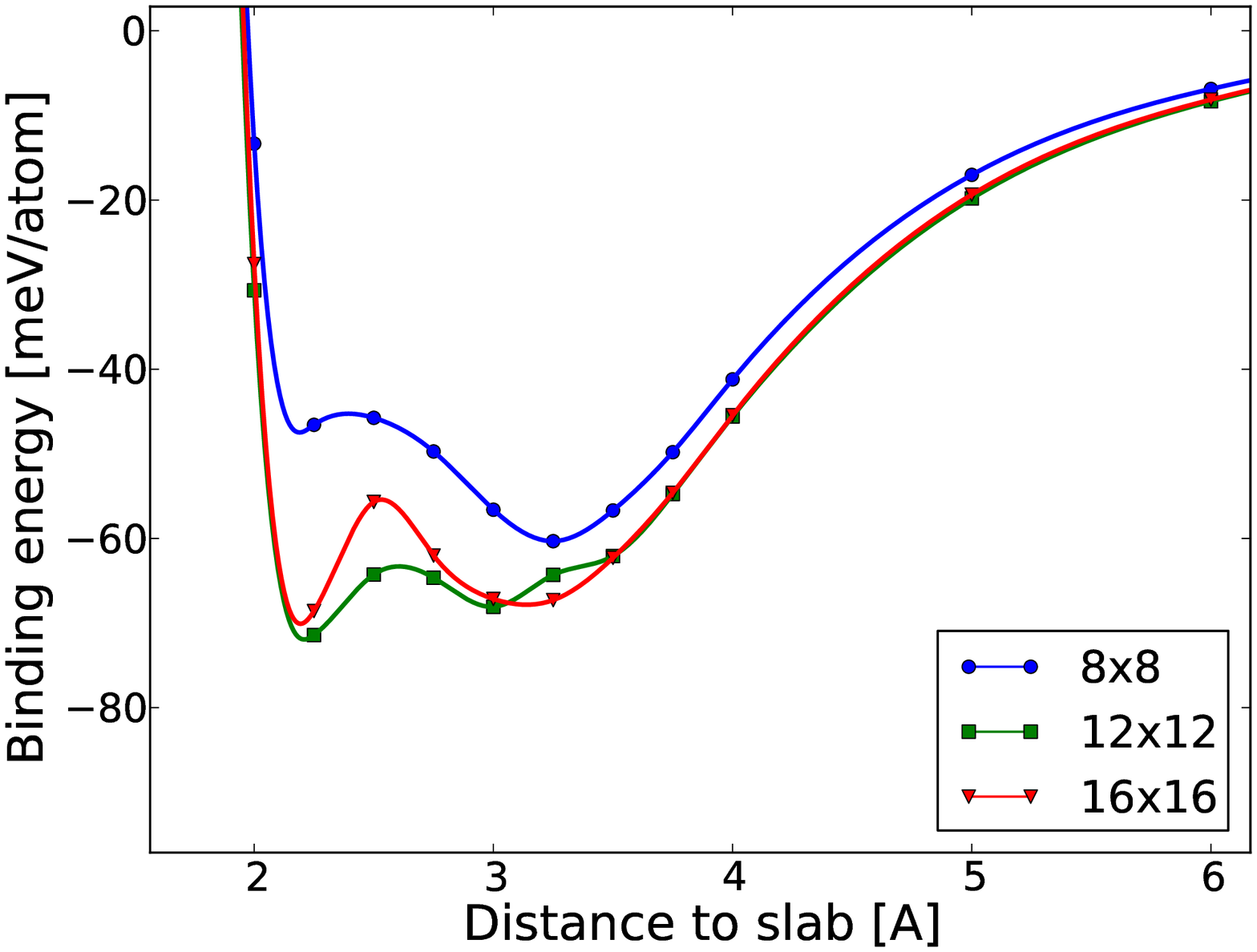}
 \includegraphics[scale=0.35]{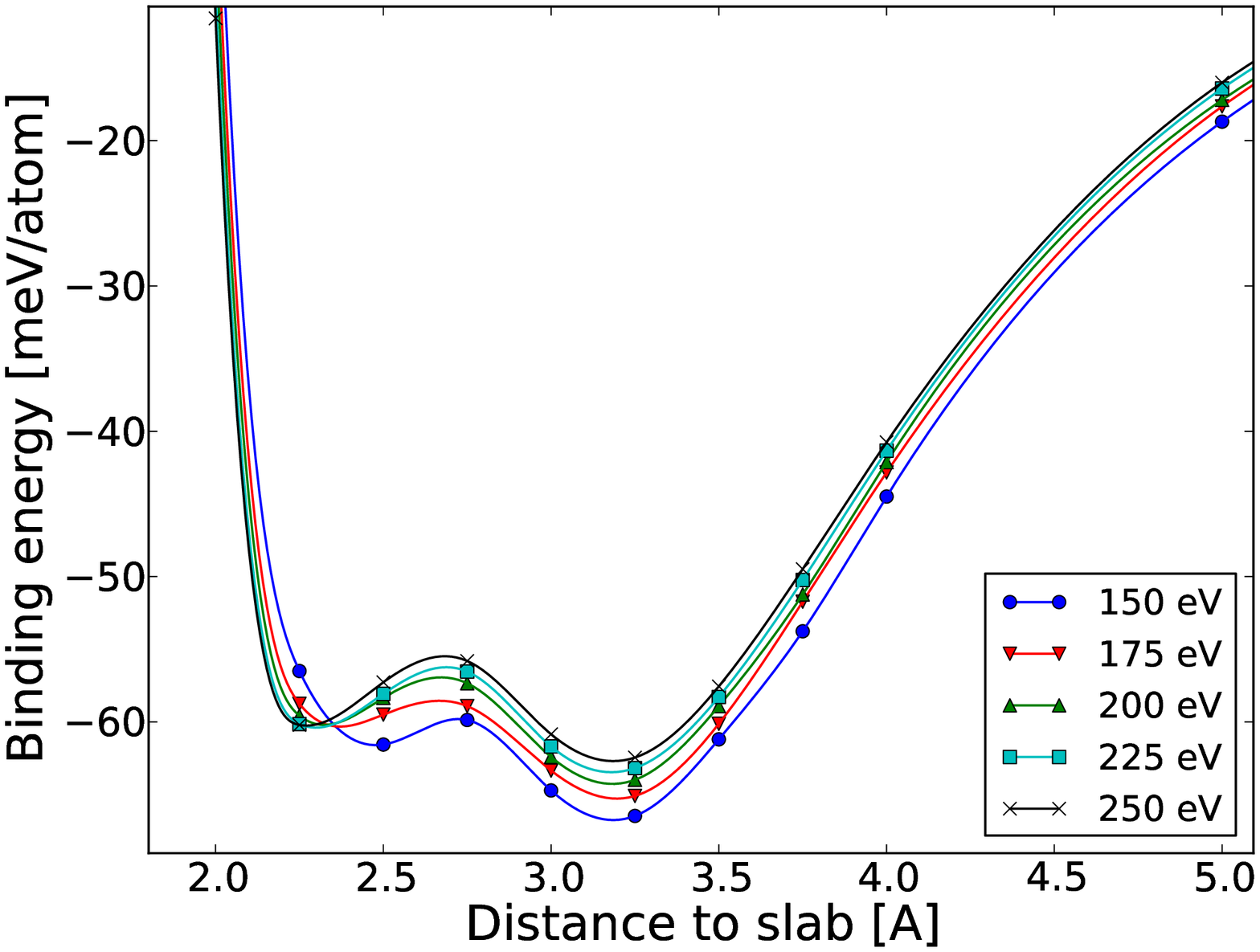}
 \caption{Energy curves for graphene on Ni(111). Upper left: HF using $8\times8$ $k$-point sampling with different Fermi smearings. Upper right: HF+RPA using $8\times8$ $k$-point sampling and $E^{RPA}_{cut}=150$ eV with different Fermi smearings. Middle left: HF at different $k$-point samplings. Middle right: HF+RPA at different $k$-point samplings using and $E^{RPA}_{cut}=200$ eV. Bottom: HF+RPA at different cutoff energies for the RPA correlation energy using $8\times8$ $k$-point sampling}
\label{graph_Ni}
\end{center}
\end{figure*}

In Fig. \ref{graph_Ni} we show various convergence test for graphene on Ni(111). The HF energy curves are seen to be highly dependent on both $k$-point sampling and Fermi smearing. In fact, it seems extremely difficult to converge the $k$-point sampling for the pure HF energy curves. Nevertheless, when the RPA energy is added the results are less sensitive to $k$-point sampling and Fermi smearing and converge much more rapidly. This behavior is most likely due to the error cancellation between the expressions \eqref{E_x} and \eqref{E_c}. In particular the non-interacting response function has been added and subtracted from Eq. \eqref{E_xc} and should be evaluated at the same $k$-point sampling in the two expressions. It should be remarked that the potential energy curve with $16\times16$ $k$-point sampling is nearly identical to the one obtained in Ref. [\onlinecite{mittendorfer}] with $19\times19$ $k$-point sampling and we regard the energy curve as converged. We note that the RPA energy curves are still more sensitive to Fermi smearing than the semi-local functionals and the van der Waals Functional, where a Fermi smearing of 0.1 eV is sufficient for a converged result. We also show the energy curves evaluated at cutoff energies in the range $150-250$ eV with a $8\times8$ $k$-point sampling. The largest change is seen when increasing the cutoff from 150 eV to 200 eV. In the present paper we have evaluated the RPA energy curves for the small unit cells (Ni, Cu, and Co) using 200 eV cutoff and the large unit cells (Pd, Pt, Au, Ag, and Al) using 150 eV cutoff. The choice of 150 eV for the large unit cells may not be quite enough for detailed convergence, but since the energy curves for these systems do not have much structure we believe that the results give the correct qualitative features of the RPA energy curves with a correct equilibrium distance and binding energy which is within 5 meV of the converged result.
 
\subsection{Atomization energies of molecules}\label{CO_convergence}
In some respects, convergence of molecular atomization energies seems somewhat simpler than correlation energies of bulk systems, since one does not have to worry about $k$-point sampling and Fermi smearing. On the other hand, convergence of unit cell size may become a major problem for a plane wave implementation. Furthermore extrapolation of cutoff energies is essential for molecular systems and it may be hard to obtain accurate results from two-point extrapolations using Eq. \eqref{fit}. In general, the accuracy of the extrapolation is increased with increasing unit cell size, since the increased number of plane waves at a given cutoff energy results in a smoother cutoff dependence.

\begin{figure*}[tb]
\begin{center}
 \includegraphics[scale=0.39]{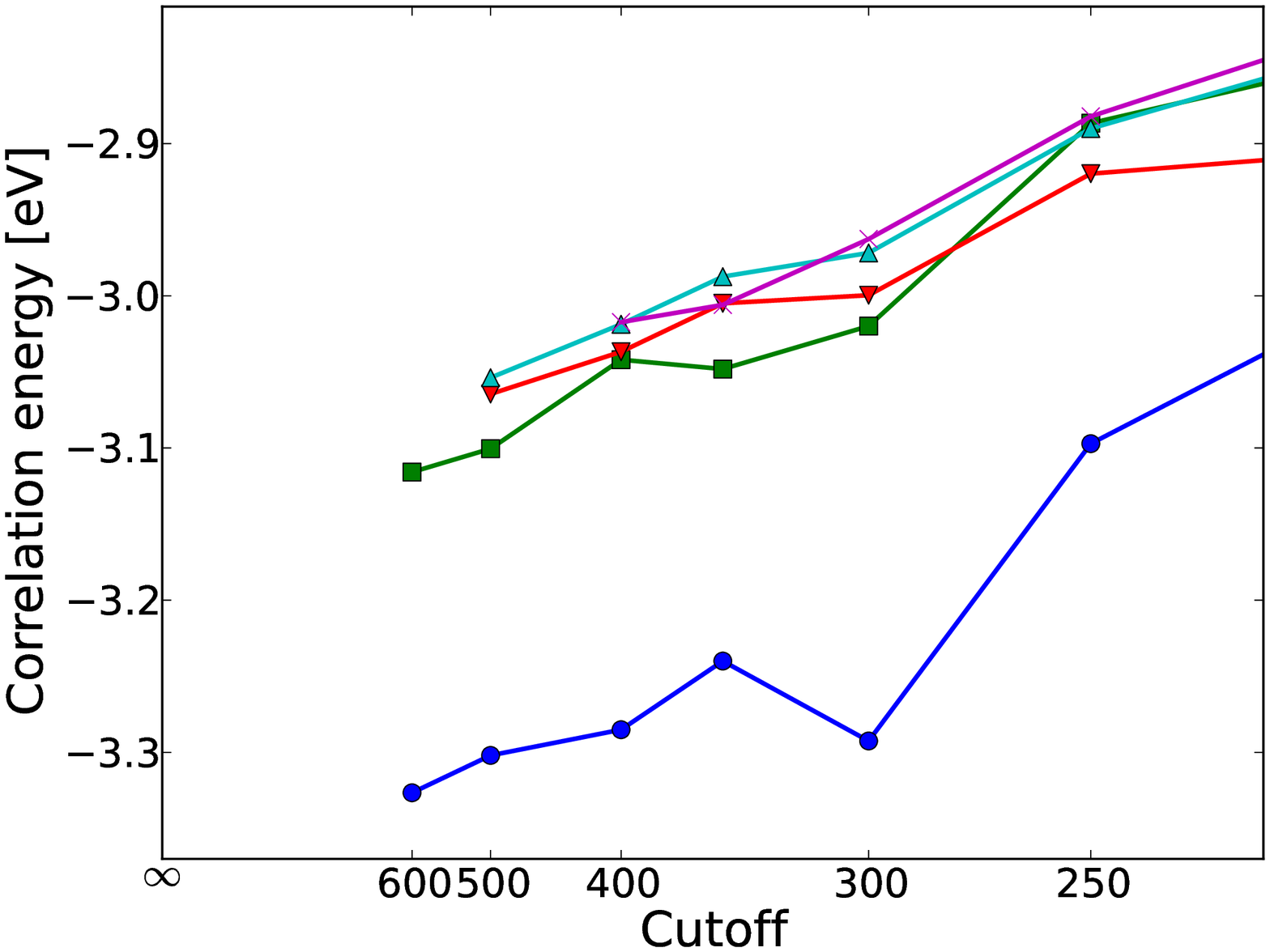}
 \includegraphics[scale=0.39]{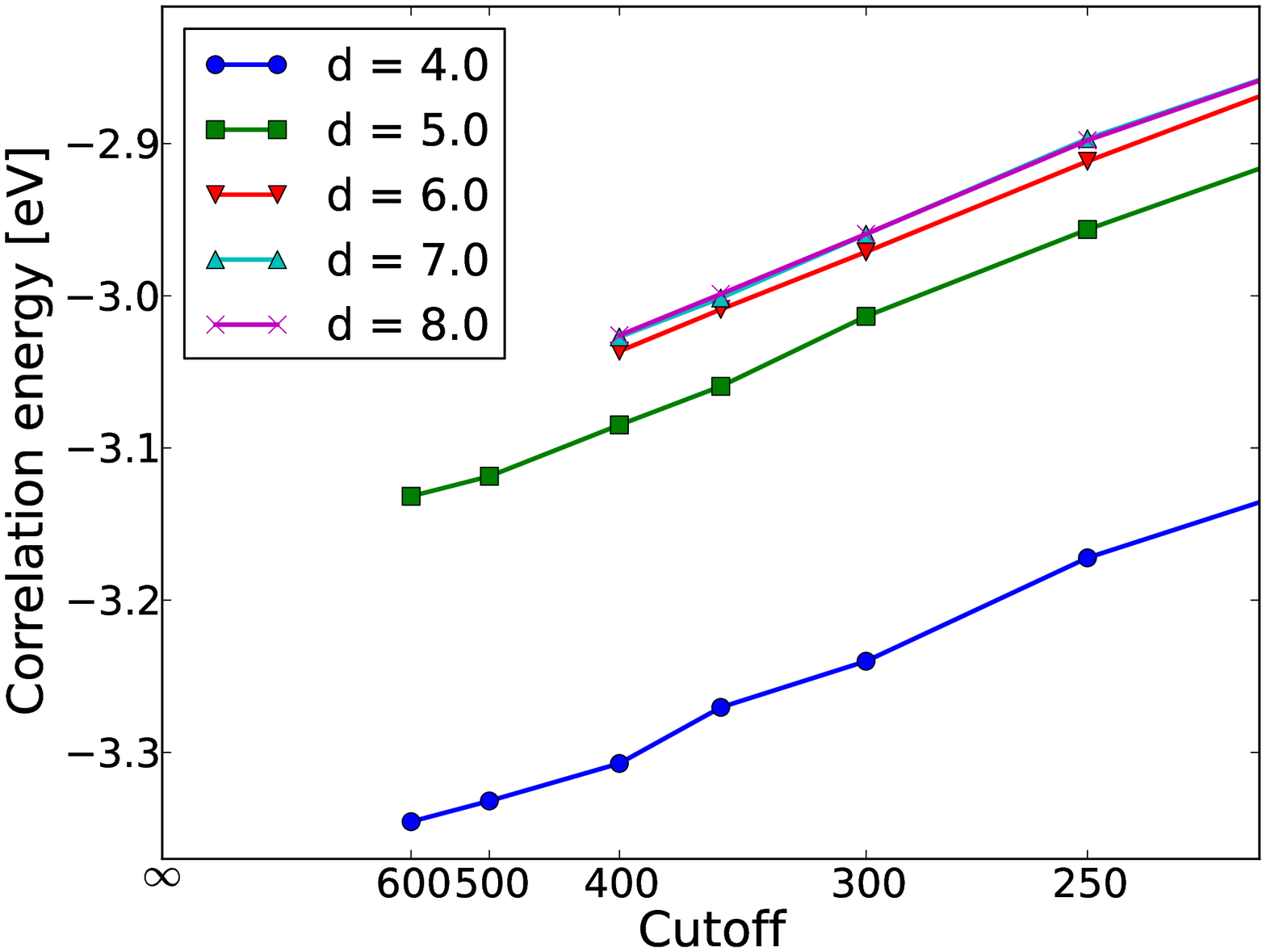}
 \includegraphics[scale=0.39]{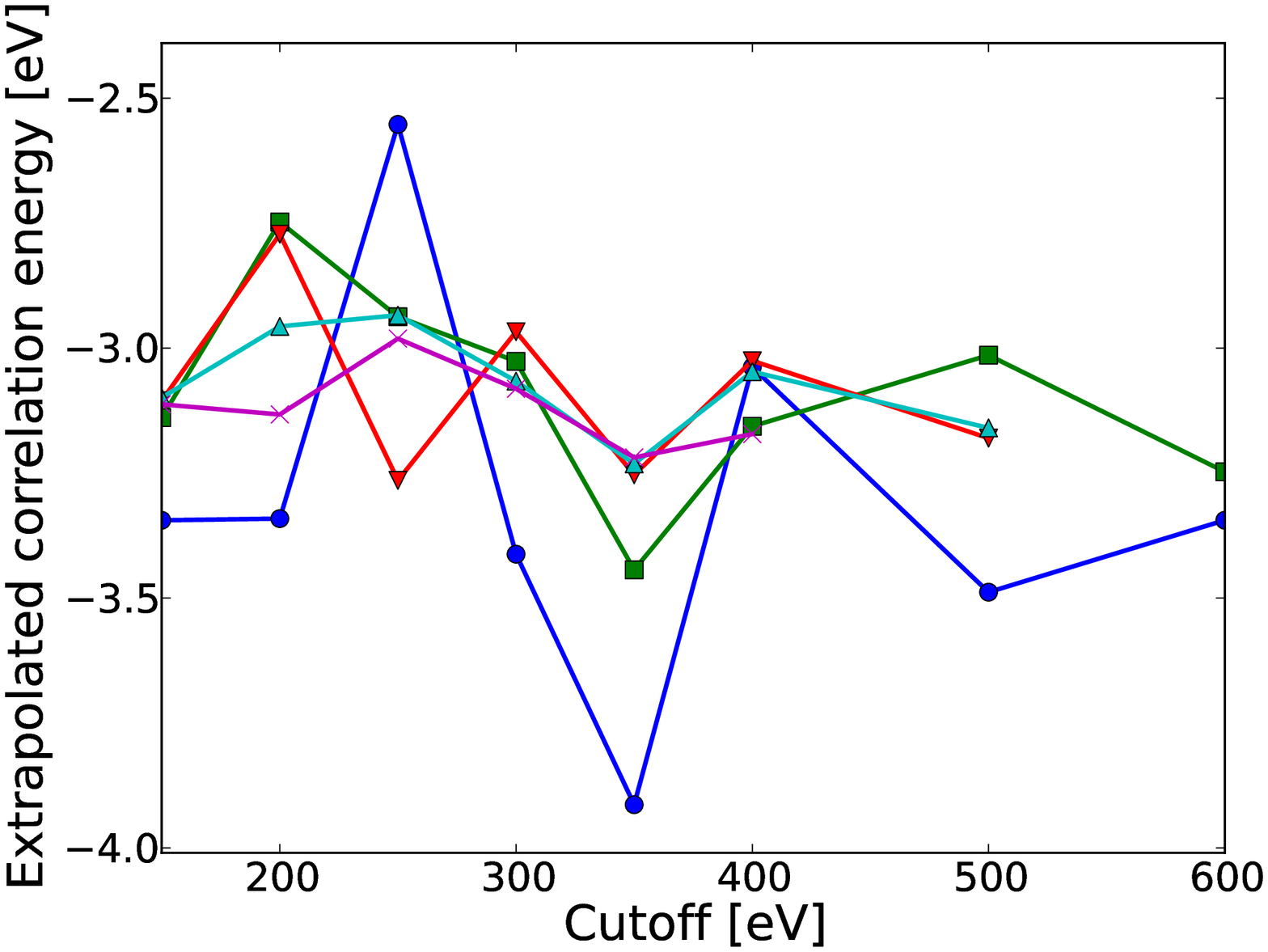}
 \includegraphics[scale=0.39]{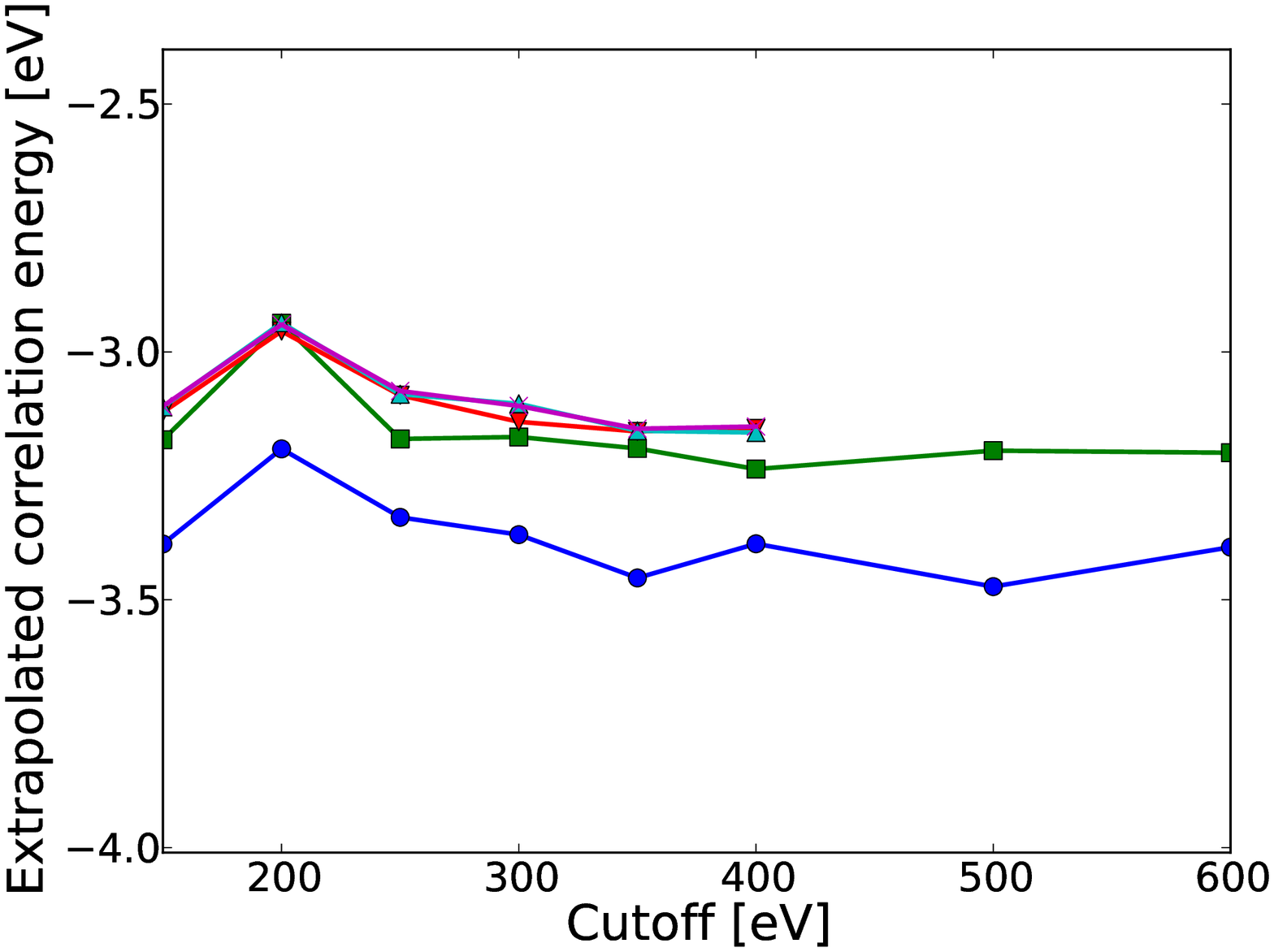}
 \caption{RPA correlation contribution to the atomization energy of CO molecule. Top: Correlation energy as a function of cutoff energy. Bottom: Extrapolated correlation energy as a function of highest extrapolation point. Left: Calculations of C, O, and CO are performed in the unit cells $(d\times d\times d)$, $(d\times d\times d)$, and $(d\times d\times d+1.1283)$ respectively. Right: Calculations of C, O, and CO are all performed in the unit cell $(d\times d\times d+1.1283)$.}
\label{CO_con}
\end{center}
\end{figure*}

Here we show a few convergence tests exemplified by the atomization energy of the CO molecule. In Fig. \ref{CO_con} we show the correlation energy contribution to the atomization energy at different unit cell sizes and the extrapolated results. The extrapolated results were obtained by a two-point extrapolation using Eq. \eqref{fit} with two subsequent cutoff points. In the left column, slightly different unit cells have been used in the evaluation of O, C, and CO correlation energies and the atomization energy converges rather roughly. In the right column, the same unit cell was used for O, C, and CO and a much smoother convergence is observed. When evaluating energy differences one can thus benefit from error cancellation when the same unit cell size and set of plane waves are used. In the present case, we see that the results are converged to within $20$ meV when the distance between periodic images exceeds $6$ {\AA}. However, for energy differences originating from different unit cell sizes, the error from the extrapolation is larger than 0.1 eV. For the same reason, it is much harder to obtain a high accuracy when evaluating the cohesive energies of solids where one cannot take energy differences between identical unit cells.

\newpage


\end{document}